\documentclass[journal,10pt,twocolumn]{IEEEtran}
\usepackage{graphicx}
\usepackage{epstopdf}
\usepackage{float}
\usepackage{amsfonts}
\usepackage{amssymb}
\usepackage{amsmath}
\usepackage{accents}
\usepackage{multirow}
\usepackage{array}
\usepackage{pgfplots}
\usepackage{booktabs}
\usepackage{multirow}
\usepackage{epstopdf,epsfig}
\usepackage{mathtools}
\usepackage{xcolor}
\DeclarePairedDelimiter{\floor}{\lfloor}{\rfloor}

\ifCLASSINFOpdf

\else

\fi

\begin{document}

\title{Tomlinson-Harashima Precoding with Stream Combiners for  MU-MIMO with Rate-Splitting \vspace{-0.05em}}
\author{A. R. Flores, R. C. de Lamare and B. Clerckx
\thanks{This work was partially supported by the National Council for Scientific and Technological Development (CNPq), FAPERJ, FAPESP, CGI, and by the U.K. Engineering and Physical Sciences Research Council (EPSRC) under
grant EP/N015312/1, EP/R511547/1.}\thanks{A. R. Flores is with the Centre for Telecommunications Studies, Pontifical
Catholic University of Rio de Janeiro, Rio de Janeiro 22451-900, Brazil (e-mail: andre.flores@cetuc.puc-rio.br).}\thanks{R. C. de Lamare is with the Centre for Telecommunications Studies,
Pontifical Catholic University of Rio de Janeiro, Rio de Janeiro 22451-900,
Brazil, and also with the Department of Electronic Engineering, University
of York, York YO10 5DD, U.K. (e-mail: delamare@cetuc.puc-rio.br).}\thanks{B. Clerckx is with the Department of Electrical and Electronic Engineering, Imperial College London, London SW7
2AZ, UK (email: b.clerckx@imperial.ac.uk).}}

\maketitle

\begin{abstract}
This paper introduces multiuser multiple-input multiple-output (MU-MIMO) architectures based on non-linear precoding and stream combining techniques using rate-splitting (RS), where the transmitter often has only partial knowledge of the channel state information (CSI). In contrast to existing works, we consider deployments where the receivers may be equipped with multiple antennas. This allows us to employ linear combining techniques based on the Min-Max, the maximum ratio and the minimum mean-square error criteria along with Tomlinson-Harashima precoders (THP) for RS-based MU-MIMO systems to enhance the sum-rate performance. Moreover, we  incorporate the Multi-Branch (MB) concept into the RS architecture to further improve the sum-rate performance. Closed-form expressions for the signal-to-interference-plus-noise ratio and the sum-rate at the receiver end are devised through statistical analysis. Simulation results show that the proposed RS-THP schemes achieve better performance than conventional linear and THP precoders.
\end{abstract}

\begin{IEEEkeywords}
Multiuser MIMO, ergodic sum-rate, rate-splitting, Tomlinson-Harashima Precoding.
\end{IEEEkeywords}

%
\IEEEpeerreviewmaketitle

\section{Introduction}
Over the last decade, considerable advances have been reported in multiple-input multiple-output (MIMO) technology. This technology achieves a high throughput by employing multiple antennas at the Base Station (BS) and at the receivers, which allows the transmission of many data streams simultaneously. MIMO has been incorporated in several wireless communications standards \cite{Li2010,Jones2015,Parkvall2017,Shafi2017,mmimo,wence,mwc} such as IEEE 802.11n, 802.16m (WiMAX), 3GPP long-term evolution (LTE), LTE-Advanced and 5G New Radio, due to their benefits. However, the simultaneous transmission of multiple streams to different users results in multiuser interference (MUI), which reduces the gains of MIMO wireless systems. 

\subsection{Prior and related work}

In order to deal with MUI, different transmit and receive processing techniques have been investigated. Specifically, the uplink of multiuser MIMO (MU-MIMO) systems requires receive processing techniques in order to take advantage of joint processing at the BS. In contrast, transmit processing techniques are employed in the downlink of MIMO systems where the receivers of multiple users do not cooperate. One of the most common techniques in the downlink involves the use of a precoder to map the symbols to the transmit antennas and form the transmitted signal. In \cite{Joham2005,rmmseprec} linear precoding techniques such as the linear mean-square error (MMSE) precoder have been developed. However, these methods suffer a substantial power loss due to the pre-processing of the transmit signal. In \cite{Spencer2004,Lai-UChoi2004,lcbd,Zu2013,wlbd} a general Zero Forcing (ZF) precoder known as Block Diagonalization (BD) has been introduced for scenarios where the receivers are equipped with multiple antennas. Later, in \cite{Stankovic2008} the Regularized Block Diagonalization (RBD) was developed to enhance the performance of BD. The main drawback of BD and RBD precoders is the high computational complexity as we scale up the number of users. In order to reduce the computational cost and improve performance of BD and RBD techniques, several algorithms have been reported in the literature such as the generalized ZF channel inversion (GZI) and the generalized MMSE channel inversion (GMI) in \cite{Sung2009} and the lattice reduction-aided simplified GMI (LR-S-GMI) in \cite{Zu2013}.

Non-linear precoding techniques achieve an overall better performance than linear precoders at the expense of a higher computational complexity. Among the non-linear precoding tehcniques, we have techniques based on discrete optimization \cite{bbprec} and Tomlinson-Harashima precoding (THP) \cite{Tomlinson1971,Harashima1972} which was originally proposed to deal with intersymbol interference (ISI) and then extended to spatial division multiple access (SDMA) in \cite{Fischer2002}. As mentioned in \cite{Windpassinger2004,Debels2018}, the THP implemented at the transmitter is the counterpart of the successive interference cancellation (SIC) technique used at the receiver \cite{spa,mfsic,dfcc,tds,mbdf,1bitidd,bfidd,listmtc,dyovers}. Similarly to SIC, the symbol order plays a critical role in the performance of THP. As a result, substantial effort has been made to develop techniques for symbol ordering. Inspired by the uplink-downlink duality, in \cite{Wolniansky1998} the decoding order has been obtained from SIC employed at the BS. In \cite{Wuebben2002,Wubben2003} low complexity ordering algorithms based on a sorted QR decomposition were proposed, reducing the computational complexity of \cite{Wolniansky1998}. A Cholesky decomposition with symmetric permutation has been proposed in \cite{Kusume2007} for optimum and suboptimum ordering. In \cite{Fung2007,Dao2010} algorithms were proposed to obtain a near optimal ordering. Because the schemes in \cite{Kusume2007,Fung2007,Dao2010} were only suitable for MU-MISO systems, a successive optimization THP (SO-THP) was proposed for MU-MIMO, attaining a BER gain over conventional THP at low SNR . The work in \cite{Zu2014} introduced multi-branch THP (MB-THP), whereas  user ordering optimization for THP was reported in \cite{Sun2015}, which can further enhance the performance of MU-MIMO systems. 

The design of the linear and non-linear precoders presented in most previous works assumes perfect knowledge of the channel state information at the transmitter (CSIT). In practice, CSIT is always obtained imperfectly \cite{locsme,okspme}, which makes the perfect CSIT assumption questionable. On the other hand, imperfect or partial CSIT results in residual interference, which degrades the overall performance of the system. In this context, rate splitting (RS) has emerged as a promising strategy to deal with the uncertainty in the CSIT \cite{Clerckx2016}. This technique splits the data into a common message and private messages. Then, both messages are superimposed, precoded and transmitted. At the receiver SIC is performed to obtain the common message, which must be decoded by all users. In contrast, the private messages are decoded only by their corresponding users after SIC. In fact, RS can adapt both the content and the power of the common stream, which allows us to adjust how much interference is treated as noise and how much interference should be decoded at the receivers, thereby increasing the flexibility of the system.

RS schemes have been shown to enhance the performance of different deployments of wireless communications systems. Furthermore, RS has been shown to achieve the optimal Degrees-of-Freedom (DoF) region when considering imperfect CSIT \cite{Piovano2017}. Moreover, RS outperforms conventional schemes such as conventional precoding in spatial division multiple access (SDMA), power-domain Non-Orthogonal Multiple Access (NOMA) \cite{Mao2018,Clerckx2020} and even Dirty Paper Coding (DPC) \cite{Mao2020}. RS has been employed in multiple-input single-output (MISO) systems along with linear precoders \cite{Joudeh2016,Hao2015} in order to maximize the sum-rate performance under perfect and imperfect CSIT assumptions. RS with non-linear precoders has been considered in \cite{Flores2018}, whose preliminary results lead us to develop this work. Another approach has been presented in \cite{Joudeh2017} where the max-min fairness problem has been studied. RS has been extended to MIMO systems in \cite{Hao2017}, where the DoF is characterized. RS has been employed in \cite{Lu2018} to mitigate the effects of the imperfect CSIT caused by finite feedback. In \cite{Flores2020}, RS with common stream combining techniques has been implemented in order to exploit the multiple antennas at the receiver and to improve the overall performance. Although previous works have shown that employing multiple antennas at the receivers improve the performance, RS in MIMO scenarios with non-linear precoders remain unexplored.

\subsection{Contributions}
In order to improve the overall rate performance of RS MU-MIMO systems, non-linear precoders can be employed at the transmit side. Since modern networks consider that the user equipment (UE) may be equipped with multiple antennas, common stream combiners at the receiver can be used to further enhance the sum-rate performance. With this approach, we generalize previous works on RS with non-linear precoders. The main contributions of this work are summarized as follows:
\begin{itemize}
\item{Non-linear precoding schemes for RS are presented, complementing the linear approaches reported in literature\cite{Joudeh2016,Mao2018}. Specifically, two Rate-Splitting  Tomlinson-Harashima Precoding (RS-THP) techniques for MU-MIMO systems are introduced, based on ZF and MMSE transmit filters. The proposed RS-THP techniques are designed with two different deployments, known as centralized and decentralized architectures. Since THP is known to be a practical strategy to approach DPC performance in the presence of perfect CSIT, the proposed RS-THP techniques can also be viewed as practical architectures to implement the Dirty Paper Coded Rate-Splitting architecture proposed in \cite{Mao2020} for multiple-antenna broadcast channels with imperfect CSIT.}
\item{ Furthermore, common stream combiners based on the Min-Max, maximum ratio combining (MRC) and minimum mean-square error combining (MMSEc) criteria are developed for RS-THP to exploit the benefits of  multiple antennas. In particular, the work in \cite{Flores2018} is extended to RS-THP and to receivers with multiple antennas.}
\item{We incorporate the multi-branch (MB) concept into the proposed RS-THP schemes, which performs improved symbol ordering and results in enhanced sum-rates.}  
\item{A statistical analysis of the sum-rates is carried out, which leads us to mathematical expressions to describe the signal-to-noise-plus-interference ratio (SINR) along with mathematical expressions for the sum-rate performance of the proposed techniques.}
\item{An analysis of the computational complexity of the proposed and existing schemes is also performed along with a simulation study of the proposed schemes against previously reported techniques.}
\end{itemize}

\subsection{Organization}
The rest of this paper is organized as follows. In Section II the system model is presented along with an overview of THP techniques. Section III describes the proposed RS-THP with common stream combiners. Section IV details the proposed stream combiners and Section V describes the achievable rate analysis. Simulation results are presented in Section VI. Finally, Section VII draws the conclusions of this work.
  
\subsection{Notation}
Matrices are denoted by uppercase boldface letters, whereas column vectors are denoted by lowercase boldface letters. The $i$th row of matrix $\mathbf{A}$ is expressed by $\mathbf{a}_{i,*}$. Scalars are denoted by standard letters. The operators $\lVert \cdot \rVert, \odot,$ and $\mathbb{E}_x\left[\cdot\right]$ stand for the Euclidean norm, the Hadamard product and the expectation operator w.r.t the random variable $x$. The superscripts $\left(\cdot\right)^{\text{T}}$, $\left(\cdot\right)^*$ and $\left(\cdot\right)^{\text{H}}$ denote the transpose, the complex conjugate and conjugate transpose of a matrix, respectively. The trace of a matrix is given by $\text{tr}\left(\cdot\right)$. The operator $\text{diag}\left(\mathbf{c}\right)$ creates a diagonal matrix with the entries of the vector $\mathbf{c}$ on the main diagonal. The cardinality of a set is given by $\text{card}\left(\cdot\right)$.

\section{System Model} 
Let us consider the downlink of a MU-MIMO system where the BS, which is equipped with $N_t$ antennas, transmits $M$ messages to $K$ users. Each User Equipment (UE) may have multiple antennas, where $N_k$ denotes the number of receive antennas at the $k$th user and $N_r=\sum_{k=1}^{K}N_k$ denotes the total number of receive antennas.  Moreover, it follows that $M\leq N_r$ and that the number of transmit antennas $N_t$ is larger than $N_r$. 

As illustrated on the left hand side of Fig. \ref{FigureTHP}, the system employs an RS scheme that for simplicity splits the message of a single user, namely the $k$th user\footnote{In a general RS scheme the messages of multiple users may be split. Assuming that the common message is entirely composed by the message of a single user constitutes a special case. The proposed schemes, however, aim at enhancing the total ergodic sum-rate. For this purpose, splitting the message of a single user is sufficient as explained in \cite{Hao2015}.}.  The message $m^{\left(\text{RS}\right)}$ intended for the $k$th user is split into a common message $m_c$ and a private message $m_k$. At this point, it is important to highlight that selecting a different user to split its message does not affect the overall performance of the system since the system performance is bounded by the worst user. This fact will be addressed with further details in the following sections. The messages $m_i$ of the other users $i\neq k$ are not split into common and private parts. After the splitting process, the messages are encoded, modulated and gathered into a vector of symbols $\mathbf{s}\in\mathbb{C}^{N_r+1}$ where the common stream is superimposed to the private streams, i.e., $\mathbf{s}=\left[s_c,\mathbf{s}_p\right]$. Remark that the system employs SIC at the receiver, which allows the decoding of the common symbol $s_c$. The vector $\mathbf{s}_p \in \mathbb{C}^{N_r}$ contains the private streams of all users and is given by $\mathbf{s}_p=\left[\mathbf{s}^{\text{T}}_1,\cdots,\mathbf{s}^{\text{T}}_K\right]^{\text{T}}$, where the vector $\mathbf{s}_k\in\mathbb{C}^{N_k}$ contains the $N_k$ private streams intended for the $k$th user. The set $\mathcal{M}_k$ contains the private streams intended for the $k$th user, where $\text{card}\left(\mathcal{M}_k\right)=N_k$.

The precoder $\mathbf{P}=\left[\mathbf{p}_c,\mathbf{P}_1,\cdots,\mathbf{P}_K\right]\in\mathbb{C}^{N_t\times\left(N_r+1\right)}$ maps the symbols to the transmit antennas. Specifically, the common precoder $\mathbf{p}_c\in\mathbb{C}^{N_t}$ maps the common symbol to the transmit antennas. In contrast, the private precoder $\mathbf{P}_k\in\mathbb{C}^{N_t\times N_k}$ maps the $N_k$ private symbols intended for the $k$th user to the transmit antennas. This model employs non-linear precoders which modify the data vector $\mathbf{s}$ through non-linear processing before mapping the data streams to the transmit antennas, originating the random vector $\mathbf{v}=\left[s_c,\mathbf{v}_1^{\text{T}},\cdots,\mathbf{v}_K^{\text{T}}\right]^{\text{T}} \in \mathbb{C}^{M+1}$, where $\mathbf{v}_k \in \mathbb{C}^{N_k}$ contains the private data intended for the $k$th user. Specifically, THP implements a feedback filter to successively cancel the interference among private symbols. A modulo operation is also applied to the private symbols, which leads to $\mathbb{E}\left[\lVert\mathbf{s}\rVert^2\right]\approx\mathbb{E}\left[\lVert\mathbf{v}\rVert^2\right]$ in order to avoid any power loss. The transmit vector is then given by
\begin{equation}
\mathbf{x}=\mathbf{P}\mathbf{v},
\end{equation}
where the vector $\mathbf{x}\in \mathbb{C}^{N_t}$. Moreover, the system satisfies a transmit power constraint given by $\mathbb{E}\left[\rVert\mathbf{x}\lVert^2\right]\leq E_{tr}$, where $E_{tr}$ represents the available power.

The transmit signal is broadcast to the UEs over the flat fading channels given by $\mathbf{H}^{\text{T}}=\left[\mathbf{H}_1^{\text{T}},\cdots,\mathbf{H}_K^{\text{T}}\right]^{\text{T}} \in \mathbb{C}^{N_r\times N_t}$. The matrix $\mathbf{H}^{\text{T}}_k \in \mathbb{C}^{N_k\times N_t}$ represents the channel between the BS and the $k$th user. Then, the received signal in a given channel use is given by

\begin{equation}
\mathbf{y}=\mathbf{H}^{\text{T}}\mathbf{P}\mathbf{v}+\mathbf{n},
\end{equation}
where $\mathbf{n}\in \mathbb{C}^{N_k}$ represents the additive noise and follows a complex Gaussian distribution, i.e., $\mathbf{n}\sim\mathcal{CN}\left(\mathbf{0},\sigma_n^2\mathbf{I}\right)$.
The signal obtained at the $k$th user is given by
\begin{equation}
\mathbf{y}_k=s_c\mathbf{H}^{\text{T}}_k\mathbf{p}_c+\sum_{i\in \mathcal{M}_k} v_i\mathbf{H}^{\text{T}}_k\mathbf{p}_i+\sum\limits_{j\notin \mathcal{M}_k} v_j\mathbf{H}^{\text{T}}_k\mathbf{p}_j+\mathbf{n}_k.\label{Received signal at user k}
\end{equation}
From \eqref{Received signal at user k} we can compute the power at the $i$th receive antenna at the $k$th UE, which is given by
\begin{equation}
\mathbb{E}\left[\lvert\left[\mathbf{y}_{k}\right]_i\rvert^2\right]=\lvert\left[\mathbf{H}_k\right]^{\text{T}}_{i}\mathbf{p}_c\rvert^2+\sum_{j=1}^{M}\lvert\left[\mathbf{H}_k\right]^{\text{T}}_{i}\mathbf{p}_j\rvert^2+\sigma_n^2,
\end{equation}
where $\left[\mathbf{y}_{k}\right]_i$ stands for the $i$th element of the vector $\mathbf{y}_{k}$ and  $\left[\mathbf{H}_k\right]_{i}$ denotes the $i$th column of the matrix $\mathbf{H}_k$.

\subsection{Imperfect CSIT model}

In general, the BS has only access to imperfect CSI. This CSI estimate is usually obtained through uplink training in Time Division Duplex (TDD) systems \cite{jidf} or via quantized feedback in Frequency Division Duplex (FDD) systems \cite{Vu2007,Love2008}. The imperfect CSIT is modeled through the error channel matrix $\tilde{\mathbf{H}}^{\text{T}}=\left[\tilde{\mathbf{H}}_1^{\text{T}},\cdots,\tilde{\mathbf{H}}_K^{\text{T}}\right]^{\text{T}} \in \mathbb{C}^{N_r\times N_t}$, which is introduced by the estimation procedure. It follows that $\mathbf{H}^{\text{T}}=\hat{\mathbf{H}}^{\text{T}}+\tilde{\mathbf{H}^{\text{T}}}$, where the channel estimate is given by $\hat{\mathbf{H}}^{\text{T}}$. 

We consider that the rows of the matrices $\hat{\mathbf{H}}^{\text{T}}$ and $\tilde{\mathbf{H}}^{\text{T}}$ are independent. Then, $\mathbb{E}\left[\mathbf{h}_{i}\mathbf{h}_{i}^{H}|\hat{\mathbf{h}}_{i}\right]=\hat{\mathbf{h}}_{i}\hat{\mathbf{h}}_{i}^H+\mathbf{R}_{e,i}$, where $\mathbf{R}_{e,i}$ is the channel error covariance matrix. Furthermore, the average error power induced $\tilde{\mathbf{h}}_{i}^{\text{T}}$ is measured by $\sigma_{e,i}^2=\mathbb{E}\left[\lVert \tilde{\mathbf{h}}_{i}^{\text{T}}\rVert^2\right]$ with $i=1,2,\cdots,N_r$. In general, the CSIT error scales with the SNR as $O\left(E_{tr}^{-\alpha}\right)$, i.e. $\sigma_{e,i}^2= O\left(E_{tr}^{-\alpha}\right)$, where $\alpha$ is the scaling factor that quantifies the CSIT quality as the SNR increases \cite{Yang2013}. We approach perfect CSIT as $\alpha \to +\infty$, which results in $\sigma_{e,1}^2, \sigma_{e,2}^2, \cdots,\sigma_{e,N_r}^2 \to 0$. On the other hand, $\alpha=0$ leads to a fixed CSIT quality across all values of SNR. Other finite values of $\alpha$ imply that the quality of CSIT improves with SNR. The value $\alpha=1$ is equivalent to perfect CSIT in the DoF sense \cite{Yang2013,Clerckx2016}. Therefore, we employ $\alpha \in \left[0,1\right]$. It is important to remark that $\alpha$ is related to several practical interpretations such as the number of quantization and feedback bits in FDD systems \cite{Hao2015}.

\subsection{Sum-Rate Performance}
At the receiver, SIC is performed to subtract the common symbol from the received signal that contains the private and common streams.\footnote{It is important to note that the SIC implemented at the receiver allows the simultaneous transmission of $M+1$ symbols .} This means that the instantaneous sum-rate of an RS MIMO system consists of two parts, the instantaneous common rate $R_{c}$, which is related to the common stream, and the instantaneous sum private rate, which is related to the private streams and is given by $R_p=\sum_{k=1}^K R_k$, where $R_k$ denotes the private rate of the $k$th user. Remark that $R_k$ is also the sum of multiple rates, since multiple streams are transmitted to the $k$th user. Assuming Gaussian signalling we have that $R_{c,k}=\log_2\left(1+\gamma_{c,k}\right)$, where $\gamma_{c,k}$ is the SINR obtained by the $k$th user when decoding the common message. Considering perfect CSIT, where the instantaneous sum-rate is achievable, we can set $R_c=\min_k R_{c,k}$. This ensure that all users decode the common message. 

{As mentioned in the previous sections the model established considers imperfect CSIT where the error is fixed or decays as $O\left(E_{tr}^{-\alpha}\right)$. Under such conditions, the instantaneous rate is not achievable. Therefore, we adopt the ergodic sum-rate (ESR) as the metric to assess the performance of the system \cite{Joudeh2016}. Once a channel estimate $\hat{\mathbf{H}}$ is obtained, the precoders are updated accordingly, obtaining an average sum-rate (ASR) per channel estimate. The ASR represents the average performance with respect to the channel errors given a channel estimate. The BS computes the ASR using imperfect instantaneous knowledge. Specifically we have that the average common rate is given by $\bar{R}_{c,k}=\mathbb{E}_{\mathbf{H}|\hat{\mathbf{H}}}\left[R_{c,k}|\hat{\mathbf{H}}\right]$. Equivalently we have that the average sum private rate is equal to $\bar{R}_p=\mathbb{E}_{\mathbf{H}|\hat{\mathbf{H}}}\left[R_p|\hat{\mathbf{H}}\right]$. The ergodic common rate at the $k$th user is given by $\mathbb{E}_{\hat{\mathbf{H}}}\left[\bar{R}_{c,k}\right]$. On the other hand, the ergodic sum private rate is equal to $\mathbb{E}_{\hat{\mathbf{H}}}\left[\bar{R}_p\right]$. Finally, the ESR of the system is given by
\begin{equation}
S_r=\min_{k\in \left[1,K\right]}\mathbb{E}_{\hat{\mathbf{H}}}\left[\bar{R}_{c,k}\right]+\mathbb{E}_{\hat{\mathbf{H}}}\left[\bar{R}_p\right].\label{Total ESR}
\end{equation}}
Note that in \eqref{Total ESR} the common rate is limited by the performance of the worst user. Then, from the ESR perspective splitting the message of any user $k$, with $k\in\left[1,K\right]$, does not affect the overall system performance.

%

\section{Proposed Rate-Splitting Tomlinson-Harashima Precoding (RS-THP)}
\begin{figure*}
\includegraphics[scale=0.35]{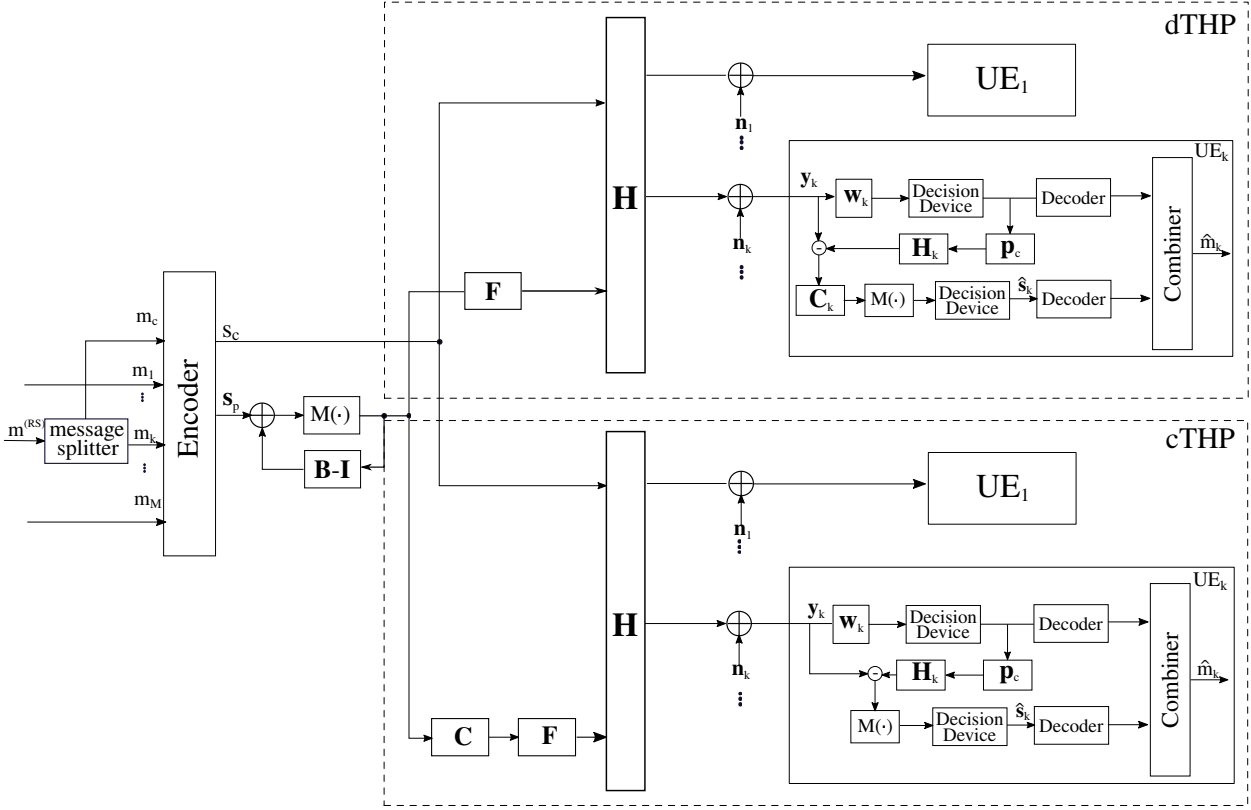}
\caption{Proposed RS-THP structures.}
\label{FigureTHP}
\end{figure*}

In this work, we propose RS-THP structures, motivated by the fact that non-linear precoding techniques outperform their linear counterparts. Furthermore, we take advantage of the multiple antennas at the UE to implement stream combiners that enhance the decoding of the common symbol. It turns out that these combiners further enhance the common rate as well. In the proposed scheme, THP acts as the private precoder of the matrix $\mathbf{P}$. We consider two different THP designs, one based on the ZF precoder and the other on the MMSE precoder. Both designs have been reported in the literature \cite{Kusume2007,Zu2014} for conventional SDMA. The MMSE design achieves a better performance than the ZF design at the expense of extra computational complexity.

Regardless of which design is considered, THP generally implements three filters. The first one is a feedback filter denoted by $\mathbf{B}\in \mathbb{C}^{N_r\times N_r}$, which is a lower triangular matrix. This matrix cancels succesively the interference caused by the previous symbols. The second filter is the feedforward filter $\mathbf{F}$, which partially removes MUI and has dimensions equal to $N_t\times N_r$ for the ZF design and $\left(N_t+N_r\right)\times N_r$ for the MMSE design. The last one is a diagonal scaling matrix given by $\mathbf{C}\in \mathbb{C}^{N_r \times N_r}$ and contains the weighted coefficients assigned to each stream. The position of the scaling matrix defines two different THP structures. The centralized THP (cTHP) implements the matrix $\mathbf{C}$ at the transmitter. In contrast, the decentralized THP (dTHP) places the scaling matrix at the receiver. Fig. \ref{FigureTHP} shows both structures considering an RS-ZF-THP scheme.


Let us first consider the MMSE-THP and define the extended channel matrix $\check{\mathbf{H}}=\left[\hat{\mathbf{H}}^{\text{T}},\sqrt{ \frac{N_r\sigma_n^2}{E_{tr}}}\mathbf{I}\right] \in \mathbb{C}^{N_r\times\left(N_r+N_t\right)}$. Computing an LQ decomposition on $\check{\mathbf{H}}$ leads to
\begin{equation}
\check{\mathbf{H}}=\check{\mathbf{L}}\check{\mathbf{Q}}=\check{\mathbf{L}}\left[\mathbf{Q}_1,\mathbf{Q}_2\right],
\end{equation}
where the unitary matrix $\check{\mathbf{Q}}$ has been split into the matrix $\mathbf{Q}_1\in\mathbb{C}^{N_r\times N_t}$ and the matrix $\mathbf{Q}_2 \in \mathbb{C}^{N_r\times N_r}$. It follows that the filters for the MMSE-THP are given by $\check{\mathbf{F}}=\check{\mathbf{Q}}^{H}$, $\check{\mathbf{C}}=\text{diag}\left(\check{l}_{1,1},\cdots,\check{l}_{N_t,N_t}\right)^{-1}$, $\check{\mathbf{B}}^{\left(\text{dTHP}\right)}=\check{\mathbf{C}}\check{\mathbf{L}}$ for the decentralized structure and $\check{\mathbf{B}}^{\left(\text{cTHP}\right)}=\check{\mathbf{L}}\check{\mathbf{C}}$ for the centralized one.

The vector $\check{\mathbf{v}}$ containing the data is successively generated as
\begin{equation}
\check{v}_i=s_i-\sum_{j=1}^{i-1}\check{b}_{i,j}\check{v}_j.\label{X feedback}
\end{equation}
From \eqref{X feedback} we know that $v_1=s_1$ and therefore $\sigma_{v_{1}}^2=\sigma_s^2$. However, for the other elements of the vector $\mathbf{x}$ we have that $\sigma_{v_{i}}\geq \sigma_s^2$. Note that the transmit power has increased due to the influence of matrix $\check{\mathbf{B}}$, originating a power loss \cite{WeiYu2005}. In order to reduce the power loss, a modulo operation is employed, which is defined element-wise as follows: 
\begin{equation}
M\left(v_i\right)=v_i-\floor[\bigg]{\frac{\text{Re}\left(v_i\right)}{\lambda}+\frac{1}{2}}\lambda-j\floor[\bigg]{\frac{\text{Im}\left(v_i\right)}{\lambda}+\frac{1}{2}}\lambda,
\end{equation}
where the coefficient $\lambda$ defines the periodic extension of the constellation and depends on the modulation alphabet and the power allocation scheme. \footnote{It is important to mention that other scenarios allow the use of different modulation order for different messages. In such cases, we employ a vector $\boldsymbol{\lambda}=[\lambda_1, \lambda_2,\cdots,\lambda_{N_r}]$ to define an appropriate modulo operation for each stream. For QAM modulations $\lambda_k=\sqrt{6M_oE_k/\left(M_o-1\right)}$, where $M_o$ represents the modulation order and $E_k$ is the power allocated to the $k$th stream.} Some common values of $\lambda$ when the symbol variance is equal to one are $\lambda=2\sqrt{2}$ and $\lambda=8/\sqrt{10}$ for QPSK and 16-QAM, respectively.

Mathematically, the feedback processing can be modeled by the inverse of the matrix $\check{\mathbf{B}}$. On the other hand, the modulo operation is equivalent to adding a perturbation vector $\mathbf{d}$ to the symbol vector $\mathbf s$ as described by 
\begin{equation}
\check{\mathbf{v}}=\check{\mathbf{B}}^{-1}\left(\mathbf{s}+\mathbf{d}\right).
\end{equation}

After precoding, the transmit vector is broadcast to the UEs through the channels. The composite signal of the users obtained at the receivers of the UEs by the proposed scheme before the SIC of the common message is described by 
\begin{align}
\mathbf{y}^{\left(\text{dTHP}\right)}=&s_c \mathbf{H}^{\text{T}}\mathbf{p}_c+\beta\mathbf{H}^{\text{T}}\mathbf{D}\check{\mathbf{F}}\check{\mathbf{v}}+\mathbf{n}\label{receive signal MMSE dthp structures}\\
\mathbf{y}^{\left(\text{cTHP}\right)}=&s_c \mathbf{H}^{\text{T}}\mathbf{p}_c+\beta\mathbf{H}^{\text{T}}\mathbf{D}\check{\mathbf{F}}\check{\mathbf{C}}\check{\mathbf{v}}+\mathbf{n},\label{receive signal MMSE cthp structures}
\end{align}
where $\mathbf{D}=\left[\mathbf{I}_{N_t},\mathbf{0}_{N_t,N_r}\right]$ and $\beta$ is the power scaling factor introduced to satisfy the transmit power constraint.

At the receivers, the common symbol is detected and then removed from \eqref{receive signal MMSE dthp structures} or \eqref{receive signal MMSE cthp structures} by performing SIC. The received signal after SIC is described by
\begin{align}
\mathbf{y}^{\left(\text{dTHP}\right)}=&\check{\mathbf{C}}\left(\beta\mathbf{H}^{\text{T}}\mathbf{D}\check{\mathbf{F}}\check{\mathbf{v}}+\mathbf{n}\right)\\
\mathbf{y}^{\left(\text{cTHP}\right)}=&\beta\mathbf{H}^{\text{T}}\mathbf{D}\check{\mathbf{F}}\check{\mathbf{C}}\check{\mathbf{v}}+\mathbf{n}.
\end{align}
Afterwards, another modulo operation is performed in order to eliminate the effects of the first modulo operation applied over the private streams. This last operation incurs in a performance penalty due the periodic extension of the constellation. In other words, some symbols at the boundary of the constellation may be mistaken by the opposite symbol. This performance degradation is known as the modulo loss. 

Let us now employ the ZF-THP, which can be obtained by directly applying the LQ decomposition to the channel matrix $\hat{\mathbf{H}}^{\text{T}}$, i.e., $\hat{\mathbf{H}}^{\text{T}}=\mathbf{L}\mathbf{Q}$. In this case, we have
\begin{align}
\mathbf{F}=&\mathbf{Q}^{H}\\
\mathbf{C}=&\text{diag}\left(l_{1,1},\cdots,l_{N_t,N_t}\right)^{-1},\\
\mathbf{B}^{\left(\text{dTHP}\right)}=&\mathbf{C}\mathbf{L},\\
\mathbf{B}^{\left(\text{cTHP}\right)}=&\mathbf{L}\mathbf{C}.
\end{align}
It follows that the received signal is given by
\begin{align}
\mathbf{y}^{\left(\textrm{dTHP}\right)}=&s_c \mathbf{H}^{\text{T}}\mathbf{p}_c+\beta \mathbf{H}^{\text{T}}\mathbf{F}\mathbf{v}+\mathbf{n}, \\
\mathbf{y}^{\left(\textrm{cTHP}\right)}=&s_c \mathbf{H}^{\text{T}}\mathbf{p}_c+\beta \mathbf{H}^{\text{T}}\mathbf{F}\mathbf{C}\mathbf{v}+\mathbf{n} .
\end{align} 
Then, the received signal for ZF-THP after removing the common message can be written as follows:
\begin{align}
\mathbf{y}^{\left(\text{dTHP}\right)}=&\beta\mathbf{C}\left(\mathbf{H}^{\text{T}}\mathbf{F}\check{\mathbf{v}}+\mathbf{n}\right),\\
\mathbf{y}^{\left(\text{cTHP}\right)}=&\beta\mathbf{H}^{\text{T}}\mathbf{F}\mathbf{C}\check{\mathbf{v}}+\mathbf{n}.
\end{align}

\subsection{Power allocation}

The proposed RS-THP schemes transmit a common stream and multiple private streams. Therefore, the available power must be allocated to both the private streams and the common stream. The power allocated to the common stream is given by $\lVert\mathbf{p}_c\rVert= \sqrt{\delta E_{tr}}$, where delta denotes the percentage of $E_{tr}$ that is designated to $s_c$ and is chosen to maximize the performance of a specific metric \cite{jpba}. It follows that the power available for the private streams is given by $E_{tr}-\lVert\mathbf{p_c}\rVert^2$ and, for simplicity, is uniformly allocated across streams. In this work, we select $\delta$ that maximizes the ESR of the system which can be mathematically expressed by
\begin{align}
\delta_o=&\max_{\delta}S_r\left(\delta\right)\nonumber\\
=&\max_{\delta}\left(\min_{k\in \left[1,K\right]}\mathbb{E}_{\hat{\mathbf{H}}}\left[\bar{R}_{c,k}\left(\delta\right)\right]+\mathbb{E}_{\hat{\mathbf{H}}}\left[\bar{R}_p\left(\delta\right)\right]\right). \label{objective power allocation delta}
\end{align}

{When $\delta=0$ no rate splitting is performed and the system is equivalent to the conventional SDMA. Since \eqref{objective power allocation delta} considers $\delta=0$ as a possible solution, the proposed schemes perform at least as well as the conventional approaches.} We remark that the power constraint must be satisfied. In this sense, the constant $\beta$ imposed to the private precoders must ensure that the transmit power constraint is fulfilled, i.e., the power of the private streams should be equal to $E_{tr}-\lVert\mathbf{p_c}\rVert^2$. For  MMSE-THP, the constant $\beta$ is given by
{\begin{align}
\beta^{\left(\textrm{cTHP}\right)}=&\sqrt{\frac{E_{tr}-\lVert \mathbf{p}_c\rVert^2}{\textrm{tr}\left( \mathbf{Q}_1\check{\mathbf{C}}\check{\mathbf{C}}^H\mathbf{Q}_1^H\right)}}\\
\beta^{\left(\textrm{dTHP}\right)}=&\sqrt{\frac{E_{tr}-\lVert \mathbf{p}_c\rVert^2}{\textrm{tr}\left( \mathbf{Q}_1\mathbf{Q}_1^H\right)}},
\end{align}}
{whereas for ZF-THP the constant takes the value of}
{\begin{align}
\beta^{\left(\textrm{cTHP}\right)}=&\sqrt{\frac{E_{tr}-\lVert \mathbf{p}_c\rVert^2}{\sum_{k=1}^{M}l_{k,k}^2}}\\
\beta^{\left(\textrm{dTHP}\right)}=&\sqrt{\frac{E_{tr}-\lVert \mathbf{p}_c\rVert^2}{M}}.
\end{align}}

\subsection{Multi-Branch THP}
Similar to the conventional THP for SDMA, the order of the symbols affects the performance of the proposed RS-THP. In this sense, incorporating a symbol ordering scheme to the proposed RS-THP structures can further improve the performance. To this end we employ a modified multi-branch (MB) processing technique. The conventional multi-branch (MB) processing was first proposed in \cite{DeLamare2008} for decision feedback receivers and further extended to THP in SDMA scenarios \cite{Zu2014}. MB-THP employs multiple parallel branches to find the best symbol ordering and enhance the performance of the THP structure. This technique generates $L_o$ different symbol ordering patterns to generate $L_o$ transmit vectors candidates. Let us consider the matrix $\mathbf{T}^{i,j}_{l}$ with $l\in 1,\cdots,L_o$, which stores one possible transmit pattern. These patterns are pre-stored at both the transmitter and the receivers. The transmitter selects, among the stored patterns, the one that leads to the highest sum-rate performance. It is important to mention that other metrics to select the optimal branch may be used.

In the MU-MIMO scenario, the patterns are designed in three steps. In the first phase, users are arranged according to several ordering patterns, which are described as follows:
\begin{align}
\mathbf{T}_{u,1}=&\mathbf{I}_K\\
\mathbf{T}_{u,i}=&\begin{bmatrix}
\mathbf{I}_{i-2} &\mathbf{0}_{i-2,K-i+2}\\
\mathbf{0}_{K-i+2,i-2} &\mathbf{\Pi}_i^{\left(u\right)}
\end{bmatrix}
,2\leq i\leq K.\label{User Order MB}
\end{align}
The matrix $\mathbf{T}_{u,i}$  in \eqref{User Order MB} denotes the $i$th ordering pattern between users. The matrix $\mathbf{\Pi}_i^{\left(u\right)}\in \mathbb{C}^{\left(K-i+2\right)\times\left(K-i+2\right)}$ exchange the order of the users and its  entries are equal to zero except on the reverse diagonal, which is filled with ones. Afterwards, for each user the streams are rearranged in a similar way, i.e.
\begin{align}
\mathbf{T}_{\mathbf{s}_k,1}=&\mathbf{I}_{N_k}\\
\mathbf{T}_{\mathbf{s}_k,j}=&\begin{bmatrix}
\mathbf{I}_{j-2} &\mathbf{0}_{j-2,N_k-j+2}\\
\mathbf{0}_{N_k-j+2,j-2} &\mathbf{\Pi}^{\left(\mathbf{s}_k\right)}_j,
\end{bmatrix}
,2\leq j\leq N_k\label{Stream Order MB}
\end{align}
where $\mathbf{T}_{\mathbf{s}_k,j}$ stands for the $j$th symbol ordering pattern of the $k$th user. In this case, the matrix $\mathbf{\Pi}_j^{\left(\mathbf{s}_k\right)}\in \mathbb{C}^{\left(N_k-j+2\right)\times\left(N_k-j+2\right)}$ exchanges the order of the symbols of the $k$th user. It is important to highlight that the matrices in \eqref{Stream Order MB} ensure that the data streams of a specific user will not be allocated to another user. We then employ both patterns $\mathbf{T}_{u,i}$ and $\mathbf{T}_{\mathbf{s}_k,j}$ together to create multiple patterns given by $\mathbf{T}_l^{\left(i,j\right)}$ with $l\in 1,\cdots,L_o$. If the users are equipped with the same number of antennas the matrix $\mathbf{T}_l^{\left(i,j\right)}$ can be found by computing the Kronecker product between $\mathbf{T}_{u,i}$ and $\mathbf{T}_{\mathbf{s}_k,j}$, as is given by
\begin{equation}
\mathbf{T}_l^{\left(i,j\right)}=\mathbf{T}_{u,i}\otimes\mathbf{T}_{\mathbf{s}_k,j}.
\end{equation}
In contrast to conventional MU-MIMO, RS schemes employs a common precoder in addition to the private precoders. Then, the best pattern is selected according to the following criterion:
\begin{align}
\mathbf{T}_{\left(o\right)}&=\max_{l}\left(\min_{k\in \left[1,K\right]}\mathbb{E}_{\mathbf{H}|\hat{\mathbf{H}}}\left[\bar{R}_{c,k}|\hat{\mathbf{H}}\left(\mathbf{T}_l\right)\right]+\mathbb{E}_{\mathbf{H}|\hat{\mathbf{H}}}\left[\bar{R}_p|\hat{\mathbf{H}}\left(\mathbf{T}_l\right)\right]\right).\label{Best patter criterion}
\end{align}

Once the best branch is selected, the channel estimate is rearranged accordingly, i.e., $\hat{\mathbf{H}}_{\left(o\right)}^{\text{T}}=\mathbf{T}_{\left(o\right)}\hat{\mathbf{H}}^{T}$. Remark that when decoding the common message, the private symbols are treated as additional noise. {Hence, the common rate is a function of $\mathbf{T}_l$}. It follows that the symbol ordering also affects the common rate and omitting the common rate in the selection criterion may lead to the selection of the wrong candidate. Moreover, setting the precoders requires $\hat{\mathbf{H}}^{\text{T}}_{\left(o\right)}$. In this sense, the computation of the common precoder should be performed by taking into account the pattern used to order the symbols. 

The proposed Multi-Branch RS-THP algorithm is summarized in Table \ref{Multi-Branch RS-THP algorithm}. The algorithm employs equation \eqref{Best patter criterion} at step 4. This criterion calculates the ASR given by $\bar{R}_{c,k}$ and $\bar{R}_p$ in order to average out the effects of the imperfect CSIT. Note that under perfect CSIT assumption the instantaneous sum-rate is achievable. In such cases the criterion given in \eqref{Best patter criterion} should use $R_{c,k}$ and $R_p$ instead of $\bar{R}_{c,k}$ and $\bar{R}_p$.

\begin{table}[ht]
\caption{}
\begin{center}
\vspace{-.3cm}
\begin{tabular}{p{0.6 cm} p{7 cm}}
\multicolumn{2}{c}{Multi-Branch RS-THP algorithm}\\
\hline
\hline
Steps & Operations \\
\hline
\rule{0pt}{3ex}
 $1$  & \bf{Form the channel matrix for the $l$th branch}\\
\rule{0pt}{3ex} 
 $2$	& \bf{Perform the LQ decomposition over the $l$th branch}\\
\rule{0pt}{1ex}
 & $\hat{\mathbf{H}}^{\left(l\right)}=\mathbf{L}^{\left(l\right)}\mathbf{Q}^{\left(l\right)}$ \\
\rule{0pt}{3ex}
$3$ &\bf{Compute the RS-THP filters for every branch}\\
	\rule{0pt}{1ex}
& $\mathbf{F}^{\left(l\right)}=\mathbf{Q}^{\left(l\right)^H}$,\\
& $ \mathbf{C}^{\left(l\right)}=\textrm{diag}\left(l_{1,1},\cdots,l_{N_t,N_t}\right)^{-1}$,\\
& $\mathbf{B}^{\left(\text{cTHP},l\right)}=\mathbf{L}^{\left(l\right)}\mathbf{C}^{\left(l\right)}$\\	
& $\mathbf{B}^{\left(\text{dTHP},l\right)}=\mathbf{C}^{\left(l\right)}\mathbf{L}^{\left(l\right)}$\\
\rule{0pt}{3ex}
$4$	&\bf{Choose the branch that maximizes the sum-rate of the RS architecture}\\
	\rule{0pt}{1ex}
 & $\mathbf{T}_{\left(o\right)}=$\\& $~~~~~~\max_{l}\left(\min\limits_{k\in \left[1,K\right]}\mathbb{E}_{\mathbf{H}|\hat{\mathbf{H}}}\left[R_{c,k}|\hat{\mathbf{H}}^{\left(l\right)}\right]+\mathbb{E}_{\mathbf{H}|\hat{\mathbf{H}}}\left[R_p|\hat{\mathbf{H}}^{\left(l\right)}\right]\right)$ \\
\rule{0pt}{3ex}
	&\bf{Perform the non-linear THP processing}\\
	\rule{0pt}{1ex} 
$5$ & $v_1=s_1$\\
\rule{0pt}{1ex}	
$6$ & \bf{for} $i=2:M$\\	
\rule{0pt}{1ex}
$7$ & ~~~~~~ $v^{\left(o\right)}_i=s_i-\sum_{j=1}^{i-1}b_{i,j}^{\left(o\right)}v^{\left(o\right)}_j$\\
\rule{0pt}{1ex}
$8$ & ~~~~~~ $v^{\left(o\right)}_i=\textrm{M}\left(v^{\left(o\right)}_i\right)$\\
\rule{0pt}{1ex}
$9$ & end \bf{for}\\
\rule{0pt}{3ex}
$10$	&\bf{Compute the transmit vector}\\
	\rule{0pt}{1ex} 
 & $\mathbf{x}^{\left(o\right)}=\mathbf{P}^{\left(o\right)}\mathbf{v}^{\left(o\right)}$\\
\rule{0pt}{3ex}
$11$	&\bf{Obtain the receive vector}\\
	\rule{0pt}{1ex} 
 & $\mathbf{y}^{\left(o\right)}=\mathbf{H}^{\left(o\right)}\mathbf{P}^{\left(o\right)}\mathbf{v}^{\left(o\right)}+\mathbf{n}$\\

\hline\label{Multi-Branch RS-THP algorithm}
\end{tabular}
\end{center}
\end{table}
\section{Stream Combining}
The proposed scheme employs RS at the transmitter to enhance the ESR. Additionally, receive processing techniques may be employed to further enhance the performance. In \cite{Ahmad2019} a MIMO cloud setting for RS with multi-antenna receivers has been studied and optimized. In \cite{Kolawole2018} the precoder and the receiver of multiuser millimeter-wave system with RS are jointly optimized to maximize the sum-rate. However, both approaches require an optimization performed for each frame, which is computationally expensive. Therefore, in \cite{Flores2020} practical stream combiners have been proposed to enhance the common rate by taking advantage of the diversity provided by the multiple antennas. These combiners improve the performance by employing the multiple antennas at the UEs. Each receiver obtains multiple copies of the common symbol leading to a better performance a system robustness. In this work, we devise stream combining techniques to support non-linear processing. 

The stream combiner $\mathbf{w}_k$ is implemented at the $k$th receiver in order take advantage of the multiple copies of the common symbol and improve the performance of the common rate. The combined signal is defined as $\tilde{y}_k=\mathbf{w}_k^H\mathbf{y}_k$. The average power of the combined signal is given by

\begin{equation}
\mathbb{E}\left[\lvert\tilde{y}_k\rvert^2\right]=\lvert\mathbf{w}_k^H\mathbf{H}_k^{\text{T}}\mathbf{p}_c\rvert^2+\sum_{j=1}^{M}\lvert\mathbf{w}_k^H\mathbf{H}_k^{\text{T}}\mathbf{q}_{b_j}\rvert^2+\lVert\mathbf{w}_k\rVert^2\sigma_n^2,\label{Combined Signal}
\end{equation}

where $\mathbf{q}_{b_j}$ depends on the THP structure employed. Let us define the matrices $\mathbf{\underline{B}}=\mathbf{B}^{-1}$ and $\check{\mathbf{\underline{B}}}=\check{\mathbf{B}}^{-1}$. Then, the ZF-THP algorithms lead to 
\begin{align}
\mathbf{q}_{b_j}^{\left(c\right)}=&\mathbf{Q}^{H}\left(\mathbf{\underline{b}}_j\odot \left[\textrm{diag}\left(\mathbf{L}\right)\right]^{-1}\right),\\
\mathbf{q}_{b_j}^{\left(d\right)}=&\mathbf{Q}^{H}\mathbf{\underline{b}}_j.
\end{align}
On the other hand, for the MMSE-THP precoders we have
\begin{align}
\mathbf{q}_{b_j}^{\left(c\right)}=&\mathbf{Q}_1^{H}\left(\check{\mathbf{\underline{b}}}_j\odot\left[\textrm{diag}\left(\mathbf{L}\right)\right]^{-1}\right),\\
\mathbf{q}_{b_j}^{\left(d\right)}=&\mathbf{Q}_1^{H}\check{\mathbf{\underline{b}}}_j.
\end{align}
  
The SINR of \eqref{Combined Signal} when decoding the common message is expressed by

\begin{equation}
\gamma_{k,c}= \frac{\lvert\mathbf{w}_k^{H}\mathbf{H}_k^{\text{T}}\mathbf{p}_c\rvert^2}{\sum\limits_{j=1}^{M}\lvert\mathbf{w}_k^H\mathbf{H}^{\text{T}}_k\mathbf{q}_{b_j}\rvert^2+\lVert\mathbf{w}_k\rVert^2\sigma_n^2}.
\end{equation}
Assuming Gaussian signalling the instantaneous common rate at the $k$th user is obtained by 
\begin{equation}
R_{c,k}=\log_2\left(1+\gamma_{c,k}\right).\label{Instantaneous common rate}
\end{equation}
In the literature, several criteria to define the combiner $\mathbf{w}_k$ have been proposed, such as the Min-Max, the MRC, and the MMSE criteria. Considering the Min-Max criterion, the common rate at the $i$th antenna of the $k$th user is given by

\begin{equation}
R_{c,k,i}=\log_2\left(1+\frac{\lvert\left[\mathbf{H}_k\right]_{i}^{\text{T}}\mathbf{p}_c\rvert^2}{\sum\limits_{j=1}^{M}\lvert\left[\mathbf{H}_k\right]_{i}^{\text{T}}\mathbf{q}_{b_j}\rvert^2+\sigma_n^2}\right).
\end{equation}
At each receiver, we select the antennas that achieve the highest ergodic common rate according to the Min-Max criterion:
\begin{equation}
R_c=\min_{k\in\left[0,K\right]}\max_{i \in \mathcal{M}_k} \mathbb{E}_{\hat{\mathbf{H}}}\left[\mathbb{E}_{\mathbf{H}|\hat{\mathbf{H}}}\left[R_{c,k,i}|\hat{\mathbf{H}}\right]\right]
\end{equation}

The MRC combiner is defined as $\mathbf{w}_k^{\left(\text{MRC}\right)}=\frac{\mathbf{H}^{\text{T}}_k\mathbf{p}_c}{\lVert\mathbf{H}^{\text{T}}_k\mathbf{p}_c\rVert^2}$. The SINR of this technique is computed by

\begin{equation}
\gamma_{c,k}^{\left(\text{MRC}\right)}=\frac{\lVert\mathbf{H}^{\text{T}}_k\mathbf{p}_c\rVert^2}{\sum\limits_{j=1}^M\lVert\mathbf{H}^{\text{T}}_k\mathbf{q}_{b_j}\rVert^2\cos^2\theta_j+\sigma_n^2},\label{SINR MRC}
\end{equation}
where $\theta_j$ denotes the angle between $\mathbf{w}_k$ and the vector $\mathbf{H}^{\text{T}}_k\mathbf{q}_{b_j}$. The instantaneous common rate can be found using \eqref{SINR MRC} in \eqref{Instantaneous common rate}.

Finally, we consider the MMSE combiner (MMSEc), which is given by
\begin{equation}
\mathbf{w}_k^{\left(\text{MMSEc}\right)}=\mathbf{R}_{\mathbf{y}_k\mathbf{y}_k}^{-1}\mathbf{H}^{\text{T}}_k\mathbf{p}_c,
\end{equation} 
and the SINR obtained with this technique is 
\begin{align}
\gamma&_{c,k}^{\left(\text{MMSEc}\right)}=\frac{\lvert\mathbf{p}_c^H\mathbf{H}_k^H\mathbf{R}_{\mathbf{y}_k\mathbf{y}_k}^{-1}\mathbf{H}_k\mathbf{p}_c\rvert}{\sum\limits_{j=1}^{M}\lvert\mathbf{p}_c^H\mathbf{H}_k^H\mathbf{R}_{\mathbf{y}_k\mathbf{y}_k}^{-1}\mathbf{H}_k\mathbf{q}_{b_j}\rvert^2+\text{tr}\left(\mathbf{R}_{\mathbf{y}_k\mathbf{y}_k}^{-2}\mathbf{H}_k\mathbf{p}_c\mathbf{p}_c^H\mathbf{H}_k^H\right)\sigma_n^2}\label{SINR MMSEc}
\end{align}
The common rate can be computed by substituting \eqref{SINR MMSEc} into \eqref{Instantaneous common rate}.

\section{Complexity and Rate Analysis}
Wireless communications system possess limited processing and hardware resources. In this sense, the computational complexity of the algorithms implemented play an essential role. Therefore, in this section, we analyze the computational complexity of the proposed and existing precoding and combining techniques. Furthermore, we derive equations that describe the achievable rates of the proposed techniques.

\subsection{Complexity Analysis}
In this section, we calculate the number of floating point operations (FLOPS) to describe the computational complexity of the proposed algorithms. Let us consider the matrices $\mathbf{Z}_1 \in \mathbb{C}^{m\times n}$ with $m<n$ and $\mathbf{Z}_2 \in \mathbb{C}^{n \times p}$.  Then, the following results hold:
\begin{itemize}
\item $\mathbf{Z}_1$ multiplied by $\mathbf{Z}_2$ requires $8mnp-2mp$ FLOPS.
\item The LQ decomposition of $\mathbf{Z}_1$ requires $8m^2\left(n-\frac{1}{3}m\right)$ FLOPS.
\end{itemize}
We consider that the LQ factorization is carried out with the Householder transformation as described in \cite{Golub1996}. Considering the special case where $m=n=p$  results in $8m^3-2m^2$ FLOPS for the multiplication of two complex matrices and $\frac{16}{3}m^3$ FLOPS for the LQ decomposition. Table \ref{ZF-THP complexity} shows the complexity of the conventional ZF-THP scheme, where we consider that $N_t=N_r=n$. Note that the computational complexity of the MMSE-THP algorithms is dominated by the LQ decomposition of the extended channel matrix, which results in $\frac{40}{3}n^3$ FLOPS. 

\begin{table}[ht]
\caption{{Computational complexity of the conventional ZF-THP with $N_t=N_r=n$}}
\begin{center}
\vspace{-.3cm}
\begin{tabular}{p{2 cm} p{3 cm} c}
\hline
\hline
Steps & Operations & FLOPS\\
\hline
\rule{0pt}{3ex} 
$1$ & LQ$\left(\mathbf{H}\right)$ & $\frac{16}{3}n^3$\\
\rule{0pt}{3ex} 
$2$ & $\mathbf{B}=\mathbf{L}\mathbf{C}$ & $n^2$\\
\rule{0pt}{3ex} 
$3$ & $\mathbf{v}$ & $4n^2+4n-8$\\
\rule{0pt}{3ex} 
$4$ & $\mathbf{F}\left(\mathbf{C'}\mathbf{v}\right)$ & $8n^2+4n$\\
\hline\label{ZF-THP complexity}
\end{tabular}
\end{center}
\end{table}

\begin{figure}[ht]
\begin{center}
\includegraphics[scale=0.48]{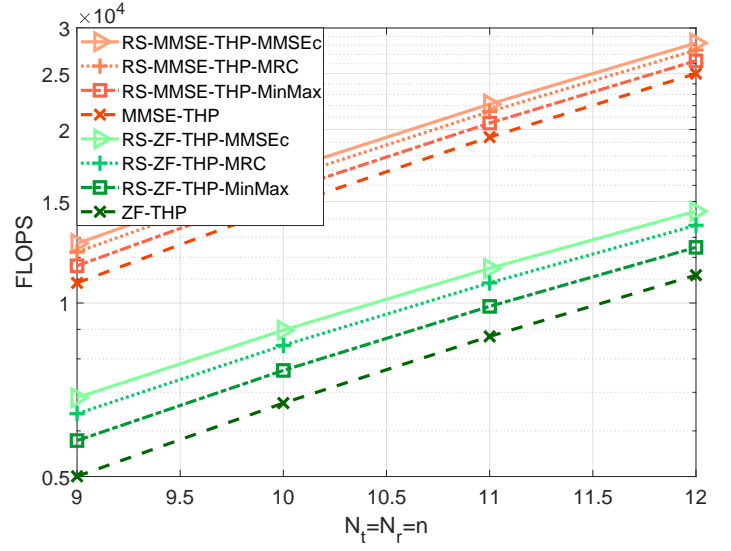}
\vspace{-1.0em}
\caption{{Complexity analysis in terms of FLOPS with $N_t=N_r=n$.}}
\label{ComputationalComplex}
\end{center}
\end{figure}

The proposed techniques include a common stream combiner which increases the performance of the system at the expense of a higher computational complexity. Table \ref{Computational complexity} summarizes the computational complexity of the proposed techniques with the use of different combiners and SIC. The required number of FLOPS of the proposed RS-THP algorithms  with stream combiners for different system dimensions is presented in Fig. \ref{ComputationalComplex}.

\begin{table}[ht]
\caption{{Computational complexity of the stream combiners with $N_t=N_r=n$}}
\begin{center}
\vspace{-.3cm}
\begin{tabular}{ p{4 cm} c}
\hline
\hline
 Technique & FLOPS\\
\hline
\rule{0pt}{3ex} 
 Min-Max & $8n-2K$\\
\rule{0pt}{3ex} 
 MRC & $8n^2+6n+6K$\\
\rule{0pt}{3ex} 
 MMSEc & $\frac{4}{3K^2}n^3+\frac{8}{K}n^2+8n^2+4n-2K$\\
\hline\label{Computational complexity stream combiners}
\end{tabular}
\end{center}
\end{table}

\begin{table}[ht]
\caption{{Computational complexity of the proposed schemes with $N_t=N_r=n$}}
\begin{center}
\vspace{-.3cm}
\begin{tabular}{ p{2.5 cm} c}
\hline
\hline
 Technique & FLOPS\\
\hline
\rule{0pt}{3ex} 
 ZF-THP & $\frac{16}{3}n^3+13n^2+8n-8$\\
\rule{0pt}{4ex} 
 RS-ZF-THP& $\frac{16}{3}n^3+21n^2+22n-2K-8$\\
 ~~~MinMax \\
\rule{0pt}{3ex} 
 RS-ZF-THP & $\frac{16}{3}n^3+29n^2+20n+6K-8$\\
 ~~~MRC\\
\rule{0pt}{3ex} 
 RS-ZF-THP & $\frac{16}{3}n^3+\frac{4}{3K^2}n^3+29n^2+\frac{8}{K}n^2+34n-2K-8$\\
 ~~~MMSEc\\
 \rule{0pt}{3ex} 
 MMSE-THP & $\frac{40}{3}n^3+13n^2+8n-8$\\
\rule{0pt}{3ex} 
 RS-MMSE-THP & $\frac{40}{3}n^3+21n^2+22n-2K-8$\\
 ~~~MinMax\\
\rule{0pt}{3ex} 
 RS-MMSE-THP & $\frac{40}{3}n^3+29n^2+20n+6K-8$\\
 ~~~MRC\\
\rule{0pt}{3ex} 
 RS-MMSE-THP & $\frac{40}{3}n^3+\frac{4}{3K^2}n^3+29n^2+\frac{8}{K}n^2+34n-2K-8$\\
 ~~~MMSEc\\
\hline\label{Computational complexity}
\end{tabular}
\end{center}
\end{table}

\subsection{Rate Analysis}

In this section, we carry out the sum-rate analysis of the proposed strategies. Furthermore, we derive expressions that describe the SINR of the proposed techniques. Assuming that the BS employs a ZF-THP with RS, the received signal at the $k$th user is given by
\begin{align}
\mathbf{y}^{\left(c\right)}_k&=s_c\mathbf{H}^{\text{T}}_k\mathbf{p}_c+\mathbf{v}_k^{\left(c\right)}+\tilde{\mathbf{H}}^{\text{T}}_k\mathbf{Q}^{H}\text{diag}\left(\mathbf{l}\right)^{-1}\mathbf{B}^{\left(c\right)^{-1}}\left(\mathbf{s}+\mathbf{d}^{\left(c\right)}\right)+\mathbf{n}_k\\
\mathbf{y}^{\left(d\right)}_k&=s_c\mathbf{H}^{\text{T}}_k\mathbf{p}_c+\mathbf{L}_k\mathbf{B}^{\left(d\right)^{-1}}\mathbf{v}^{\left(d\right)}+\tilde{\mathbf{H}}^{\text{T}}_k\mathbf{Q}^{H}\mathbf{B}^{\left(d\right)^{-1}}\left(\mathbf{s}+\mathbf{d}^{\left(d\right)}\right)+\mathbf{n}_k.\label{Received signal ZF-THP both analysis}
\end{align} 
From \eqref{Received signal ZF-THP both analysis} we can obtain the average power of the received signal at the $i$th antenna, which is given by
\begin{align}
\mathbb{E}&\left[\big\lvert y_{i}^{\left(c\right)}\big\rvert^2\right] =\Big\lvert\left(\hat{\mathbf{h}}^{\text{T}}_{i}+\tilde{\mathbf{h}}_{i}^{\text{T}}\right)\mathbf{p}_c\Big\rvert^2+\sigma_{v_i}^2+\sum_{j=1}^{N_r}\frac{1}{\lvert l_{j,j}\rvert^2}\big\lvert\tilde{\mathbf{h}}_{i}^{\text{T}}\mathbf{q}^H_{j,*}\big\rvert^2+\sigma_n^2,\label{mean power ZFdTHP per antenna}
\end{align}
\begin{align}
\mathbb{E}&\left[\big\lvert y_{i}^{\left(d\right)}\big\rvert^2\right]=\Big\lvert\left(\hat{\mathbf{h}}_{i}^{\text{T}}+\tilde{\mathbf{h}}_{i}^{\text{T}}\right)\mathbf{p}_c\Big\rvert^2+\big\lVert\left[\mathbf{L}^{\text{T}}\right]_{i}\big\rVert^2 +\sum_{j=1}^{N_r}\big\lvert\tilde{\mathbf{h}}^{\text{T}}_{i}\mathbf{q}^H_{j,*}\big\rvert^2+\sigma_n^2.\label{mean power ZFcTHP per antenna}
\end{align}
The SINR of the proposed techniques can be computed with \eqref{mean power ZFcTHP per antenna} and \eqref{mean power ZFdTHP per antenna}, which lead us to

\begin{align}
\gamma_{c,k,i}^{\left(c\right)}&=\frac{\lvert\hat{\mathbf{h}}_{i}^{\text{T}}\mathbf{p}_c\rvert^2}{\lvert\tilde{\mathbf{h}}^{\text{T}}_{i}\mathbf{p}_c\rvert^2+\sigma_{v_{i}}^2+\sum\limits_{j=1}^{N_r}\frac{1}{\lvert l_{j,j}\rvert^2}\lvert\tilde{\mathbf{h}}^{\text{T}}_{i}\mathbf{q}^H_{j,*}\rvert^2+\sigma_n^2}\label{SINR when decoding common rate analysis 1}\\
\gamma_{c,k,i}^{\left(d\right)}&=\frac{\lvert\hat{\mathbf{h}}_{i}^{\text{T}}\mathbf{p}_c\rvert^2}{\lvert\tilde{\mathbf{h}}^{\text{T}}_{i}\mathbf{p}_c\rvert^2+\big\lVert\left[\mathbf{L}^{\text{T}}\right]_{i}\big\rVert^2 +\sum\limits_{j=1}^{N_r}\lvert\tilde{\mathbf{h}}^{\text{T}}_{i}\mathbf{q}^H_{j,*}\rvert^2+\sigma_n^2}\label{SINR when decoding common rate analysis 2}
\end{align}
We can get the common rate of the RS-ZF-cTHP and RS-ZF-dTHP structures by using \eqref{SINR when decoding common rate analysis 1} and \eqref{SINR when decoding common rate analysis 2}, respectively. First, we compute the ASR, given by $\mathbb{E}\left[\log_2\left(1+\gamma_{c,k,i}\right)|\hat{\mathbf{H}}\right]$, to select at the $k$th user the antenna that leads to the best performance, i.e. $i^{\left(o\right)}_k=\max\limits_{i\in\mathcal{M}_k}\mathbb{E}\left[\log_2\left(1+\gamma_{c,k,i}\right)|\hat{\mathbf{H}}\right]$. Finally, for the Min-Max criterion, we set $\mathbf{w}_k=\mathbf{e}_{i_k^{\left(o\right)}}$, where  $\mathbf{e}_i$ is the column index vector, which contains an entry equal to one at the $i$th position and zeros at any other position.  

For the MRC combiner we need to find the value of $\lVert\mathbf{H}^{\text{T}}_k\mathbf{q}_{b_i} \rVert^2$ with $i= 1,2,\cdots, M$, which corresponds to the the multiplication of the $k$th user channel with the private precoders. In order to find this value, we need to expand the terms of the products given by $\mathbf{H}^{\text{T}}_k\mathbf{Q}^{H}\left(\text{diag}\left(\mathbf{l}\right)\right)^{-1}\mathbf{B}^{\left(\text{c}\right)^{-1}}$ and $\mathbf{H}^{\text{T}}_k\mathbf{Q}^{H}\mathbf{B}^{\left(\text{d}\right)^{-1}}$. Then, we have
\begin{align}
\mathbf{H}^{\text{T}}_k\mathbf{Q}^{H}\left(\text{diag}\left(\mathbf{l}\right)\right)^{-1}\mathbf{B}^{\left(\text{c}\right)^{-1}}&=\mathbf{H}^{\text{T}}_k\left[\frac{1}{l_{1,1}}\mathbf{q}^H_{1,*} \quad \frac{1}{l_{2,2}}\mathbf{q}^H_{2,*} \quad \cdots \quad \frac{1}{l_{M,M}}\mathbf{q}^H_{M,*}\right]\mathbf{B}^{\left(\text{c}\right)^{-1}},\label{HPcTHP Extense}\\
\mathbf{H}^{\text{T}}_k\mathbf{Q}^{H}\mathbf{B}^{\left(\text{d}\right)^{-1}}&=\mathbf{H}^{\text{T}}_k\left[\mathbf{q}^H_{1,*} \quad \mathbf{q}^H_{2,*} \quad \cdots \quad \mathbf{q}^H_{M,*}\right]\mathbf{B}^{\left(\text{d}\right)^{-1}}.\label{HPdTHP Extense}
\end{align}
Using \eqref{HPcTHP Extense} and \eqref{HPdTHP Extense} we can obtain the vector $\mathbf{H}^{\text{T}}_k\mathbf{q}_{b_i}$, which is given by
\begin{equation}
\mathbf{H}^{\text{T}}_k\mathbf{q}^{\left(\text{c}\right)}_{b_i}=
\begin{bmatrix}
h_{1,1}^{\left(k\right)} \sum\limits_{j=1}^M\frac{1}{l_{jj}}q_{j,1}\underline{b}_{j,i}^{\left(\text{c}\right)}+\cdots+h_{1,M}^{\left(k\right)} \sum\limits_{j=1}^M\frac{1}{l_{jj}}q_{j,M}\underline{b}_{j,i}^{\left(\text{c}\right)}\\
h_{2,1}^{\left(k\right)} \sum\limits_{j=1}^M\frac{1}{l_{jj}}q_{j,1}\underline{b}_{j,i}^{\left(\text{c}\right)}+\cdots+h_{2,M}^{\left(k\right)} \sum\limits_{j=1}^M\frac{1}{l_{jj}}q_{j,M}\underline{b}_{j,i}^{\left(\text{c}\right)}\\
\vdots\\
h_{N_k,1}^{\left(k\right)} \sum\limits_{j=1}^M\frac{1}{l_{jj}}q_{j,1}\underline{b}_{j,i}^{\left(\text{c}\right)}+\cdots+h_{N_k,M}^{\left(k\right)} \sum\limits_{j=1}^M\frac{1}{l_{jj}}q_{j,M}\underline{b}_{j,i}^{\left(\text{c}\right)}
\end{bmatrix},
\end{equation}
\begin{equation}
\mathbf{H}^{\text{T}}_k\mathbf{q}^{\left(\text{d}\right)}_{b_i}=
\begin{bmatrix}
h_{1,1}^{\left(k\right)} \sum\limits_{j=1}^M q_{j,1}\underline{b}^{\left(\text{d}\right)}_{j,i}+\cdots+h_{1,M}^{\left(k\right)} \sum\limits_{j=1}^M q_{j,M}\underline{b}_{j,i}^{\left(\text{d}\right)}\\
h_{2,1}^{\left(k\right)} \sum\limits_{j=1}^Mq_{j,1}\underline{b}_{j,i}^{\left(\text{d}\right)}+\cdots+h_{2,M}^{\left(k\right)} \sum\limits_{j=1}^M q_{j,M}\underline{b}_{j,i}^{\left(\text{d}\right)}\\
\vdots\\
h_{N_k,1}^{\left(k\right)} \sum\limits_{j=1}^M q_{j,1}b_{j,i}^{\left(\text{d}\right)}+\cdots+h_{N_k,M}^{\left(k\right)} \sum\limits_{j=1}^M q_{j,M}\underline{b}_{j,i}^{\left(\text{d}\right)}
\end{bmatrix}.
\end{equation}
Finally, we calculate the value of $\lVert\mathbf{H}^{\text{T}}_k\mathbf{p}_i\rVert^2$ as follows
\begin{equation}
\lVert\mathbf{H}^{\text{T}}_k\mathbf{q}_{b_i}^{\left(\text{c}\right)}\rVert^2=\sum_{p=1}^{N_k}\bigg\lvert \sum_{n=1}^{M}\sum_{j=1}^{M}\frac{1}{l_{j,j}}h_{p,n}^{\left(k\right)}q_{j,n}\underline{b}_{j,i}^{\left(\text{c}\right)}\bigg\rvert^2,\label{cthp MRC element calc}
\end{equation}

\begin{equation}
\lVert\mathbf{H}^{\text{T}}_k\mathbf{q}_{b_i}^{\left(\text{d}\right)}\rVert^2=\sum_{p=1}^{N_k}\bigg\lvert \sum_{n=1}^{M}\sum_{j=1}^{M}h_{p,n}^{\left(k\right)}q_{j,n}\underline{b}_{j,i}^{\left(\text{d}\right)}\bigg\rvert^2\label{dthp MRC element calc}
\end{equation}
By substituting \eqref{cthp MRC element calc} and \eqref{dthp MRC element calc} in \eqref{SINR MRC} we obtain the SINR of the RS-THP-MRC technique.

Note that the power loss, which depends on the modulation, was neglected in this analysis. To include the power loss, a factor $\tau$ should be introduced so that $\mathbb{E}\left[\mathbf{v}\mathbf{v}^H\right]=\tau^{-1}\mathbf{I}$ as in \cite{Flores2018}. 

\section{Simulations}

In this section, we assess the sum-rate performance of the proposed RS-THP structures under imperfect and perfect CSIT. We consider that the BS is equipped with $N_t=12$ transmit antennas and broadcasts data to $K=6$ users, each equipped with $N_k=2$ receive antennas. For conventional precoding schemes a total of 12 data streams is transmitted whereas RS schemes include additionally the common stream. The channel matrix $\mathbf{H}^{\text{T}}$ is assumed to be a complex i.i.d Gaussian matrix where each entry has zero mean and unit variance, i.e., $\mathcal{CN}\left(0,1\right)$. The entries of $\tilde{\mathbf{H}}^{\text{T}}$ are i.i.d and follow a Gaussian distribution, i.e., $\mathcal{CN}\left(0,\sigma_e^2\right)$. At the receiver white Gaussian noise with variance equal to $\sigma_n^2=1$ is added to the transmitted signal and the SNR is given by $E_{tr}$.  The inputs are Gaussian distributed with zero mean and unit variance. The ergodic sum-rate was computed considering 100 independent channel realizations. For each channel realization we consider 100 different error matrices. We perform an SVD over the channel matrix, i.e., $\mathbf{H}=\mathbf{U\mathbf{\Psi}\Upsilon}$ and set the common precoder to $\mathbf{p}_c=\boldsymbol{\upsilon}_1$.

\begin{figure}[h]
\begin{center}
\includegraphics[scale=0.45]{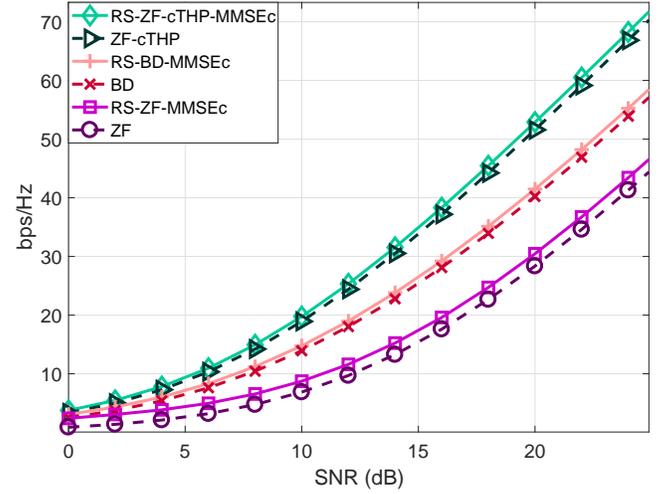}
\vspace{-1.0em}
\caption{{Sum-rate performance of a MU-MIMO system employing different ZF schemes with Perfect CSIT, $K=6$, $N_k=2$, $N_t=12$ and $\sigma_n^2=1$.}}
\label{Figure3}
\end{center}
\end{figure}

In the first example, we consider RS with several ZF precoding schemes under perfect CSIT. Fig. \ref{Figure3} summarizes the sum-rate performance achieved .Note that the multiple antennas at the UE allow us to implement the common combiners and enhance the common rate. In contrast, for the private rate of the ZF techniques each receive antenna may be treated as a separate single-antenna user. The inclusion of the terms RS and MMSEc in the legend denote that the schemes employ RS with an MMSE combiner. Conventional precoding schemes omit these terms. Note that each stream transmitted provides a gain from ZF-THP over the conventional ZF precoder, which results in a substantial ESR gain. Although the major contribution of RS is to cope with CSIT uncertainties, Fig. \ref{Figure3} shows that the RS with the MMSEc combiner at the receiver provides a consistent gain of at least one bit/Hz for all the schemes examined. Therefore, RS-THP is a promising strategy even under perfect CSIT assumption. The benefits of RS are greater under imperfect CSIT, as shown in the following simulations. 

\begin{figure}[h]
\begin{center}
\includegraphics[scale=0.48]{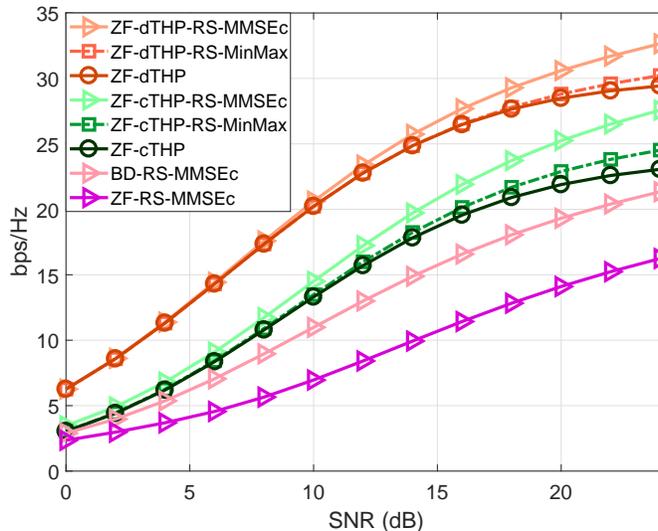}
\vspace{-1.0em}
\caption{{Sum-rate performance of a MU-MIMO system employing ZF-based precoding schemes with $\sigma_e^2=0.05$, $K=6$, $N_k=2$, $N_t=12$ and $\sigma_n^2=1$.}}
\label{Figure4}
\end{center}
\end{figure}

\begin{figure}[h]
\begin{center}
\includegraphics[scale=0.45]{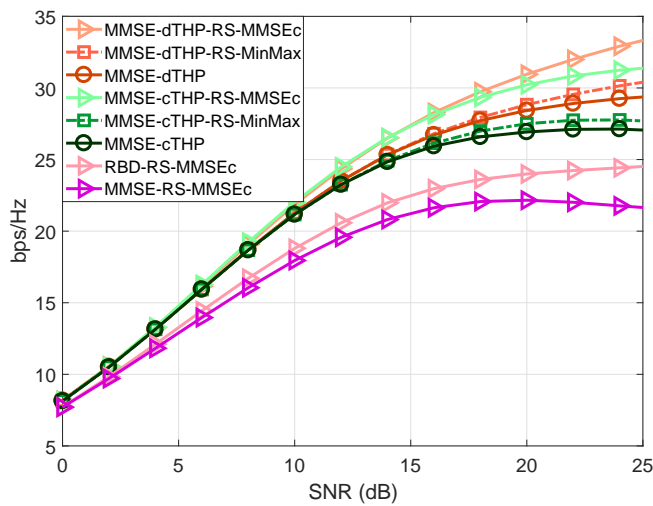}
\vspace{-1.0em}
\caption{{Sum-rate performance of a MU-MIMO system employing MMSE-based precoding schemes with $\sigma_e^2=0.05$, $K=6$, $N_k=2$, $N_t=12$ and $\sigma_n^2=1$.}}
\label{Figure5}
\end{center}
\end{figure}

In the second example, we explore an imperfect CSIT scenario where we set the variance of the error to $\sigma_e^2=0.05$. Fig. \ref{Figure4} shows the sum-rate performance obtained by the ZF-cTHP and the ZF-dTHP structures, whereas Fig. \ref{Figure5} shows the MMSE-cTHP and MMSE-dTHP structures. From Fig. \ref{Figure4} we notice that all dTHP schemes outperform those using the cTHP architecture. However, this dTHP schemes require more complex receivers. {The proposed RS-THP schemes depicted in Fig. \ref{Figure4} outperform their conventional ZF-cTHP and ZF-dTHP counterparts as expected.} Fig. \ref{Figure5} shows that MMSE-THP achieves better performance than that of ZF-THP for all the combiners employed, as expected especially for SNR values below 15 dB. This gain comes from the reduced interference due to the MMSE precoders, which results in a reduction of the power loss. The results show that RS-THP attains better performance than other schemes under CSIT uncertainties. The common combiners implemented at the receiver further increase the RS-THP performance, while MMSEc is the best combiner. 

\begin{figure}[h]
\begin{center}
\includegraphics[scale=0.45]{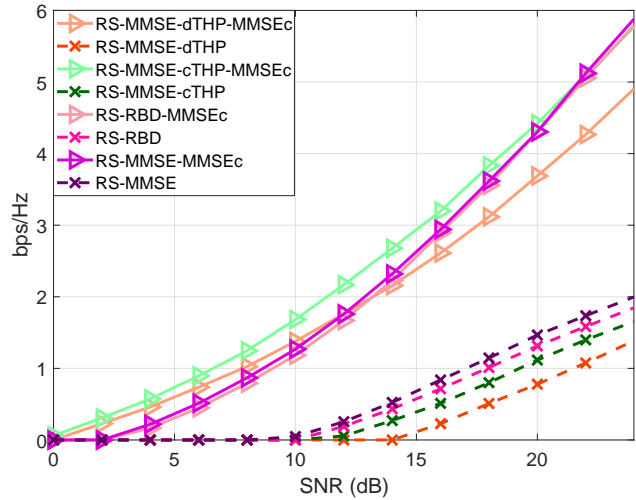}
\vspace{-1.0em}
\caption{{Common rate performance of a MU-MIMO system employing MMSE-based precoding schemes with $\sigma_e^2=0.05$, $K=6$, $N_k=2$, $N_t=12$ and $\sigma_n^2=1$.}}
\label{Figure6}
\end{center}
\end{figure}

In the third example, we assess the common rate attained by the combiners. The variance of the error was set to  $\sigma_e^2=0.05$, as in the previous experiment. From Fig. \ref{Figure6} we can see that the common combiners greatly increase the performance of the common rate when compared to conventional RS. Note that linear precoders without combiners achieve a higher common rate than non-linear ones since they saturate faster and allocate more power to the common stream in order to deal with CSIT uncertainties. 

\begin{figure}[h]
\begin{center}
\includegraphics[scale=0.45]{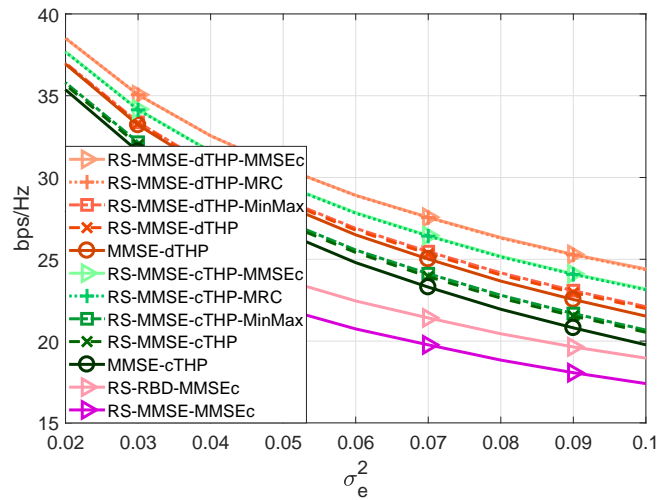}
\vspace{-1.0em}
\caption{{Sum-rate performance vs error variance at an SNR equal to 20 dB considering a MU-MIMO system employing MMSE-based precoders with $K=6$, $N_k=2$, $N_t=12$ and $\sigma_n^2=1$.}}
\label{Figure7}
\end{center}
\end{figure}

In the fourth example, we consider that the SNR is fixed at $20$dB. Table \ref{Error Var Performance Table} summarizes the ESR performance of the ZF-based schemes at different levels of $\sigma_e^2$. Remark that the ZF-cTHP obtains a gain of approximately 11 bps/Hz at $\sigma_e^2=0.05$ when compared to the conventional ZF technique. However, as shown in Table \ref{Error Var Performance Table}, this gain decays as the variance of the error increases, reducing the gap between the ZF-THP schemes and linear ZF precoder. On the other hand, Fig. \ref{Figure7} presents the ESR obtained by the MMSE-based schemes. The quality of the CSIT estimate, i.e. $\sigma_e^2$, varies in $ \left[0.02, 0.1\right]$. The MMSE-dTHP structures perform better than their corresponding MMSE-cTHP structures. From the results we can conclude that the RS-MMSE-dTHP-MMSEc scheme is the most robust strategy among the proposed ones. As expected, multiple antennas at the receiver increase the robustness of the system since the combiners allow the receivers to process multiple copies of the common symbol. Moreover, the robustness increases with the number of antennas at the receiver. 

\begin{figure}[h]
\begin{center}
\includegraphics[scale=0.45]{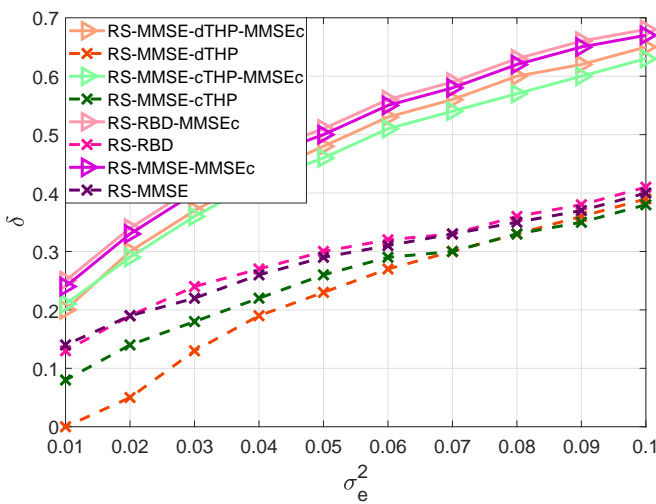}
\vspace{-1.0em}
\caption{{Power allocated to the common stream vs error variance at an SNR equal to 20 dB considering a MU-MIMO system employing MMSE-based precoders with $K=6$, $N_k=2$, $N_t=12$ and $\sigma_n^2=1$.}}
\label{Figure8}
\end{center}
\end{figure}

In Fig. \ref{Figure8}, the power allocated by each technique to the common stream is shown. We can see that the power allocated to the common stream increases as the power of the error gets higher. Linear precoders allocate more power to the common stream than non-linear techniques in order to deal with CSIT uncertainties since they saturate faster. 


\begin{table}[ht]\small
\begin{center}
\caption{{ESR of a MU-MIMO system employing ZF-based precoders with $K=6$, $Ni=2$, $N_t=12$ and different $\sigma_e^2$}}
\begin{tabular}{c c c c}
\hline
\hline
\multicolumn{1}{ c|  }{\multirow{2}{*}{Scheme}}    &\multicolumn{3}{c}{ \rule{0pt}{3ex} Error Variance}\\ 
 
\multicolumn{1}{ c| }{}     &\multicolumn{1}{c}{\rule{0pt}{3ex} $\sigma_e^2=0.05$} &$\sigma_e^2=0.1$ &\multicolumn{1}{c}{$\sigma_e^2=0.2$} \\
 \hline
 \multicolumn{1}{ c }{ZF} \rule{0pt}{3ex} &$9.88$ &$6.56$ &$3.90$ \\

  \multicolumn{1}{ c}{ZF-cTHP} \rule{0pt}{3ex} &$21.62$ &$15.43$ &$9.84$ \\

   \multicolumn{1}{ c}{ZF-dTHP} \rule{0pt}{3ex} &$28.21$ &$21.45$ &$14.78$ \\

 \multicolumn{1}{ c }{RS-ZF-MMSEc} \rule{0pt}{3ex} &$14.22$ &$11.55$ &$9.30$ \\

  \multicolumn{1}{ c}{RS-ZF-cTHP-MMSEc} \rule{0pt}{3ex} &$25.16$ &$19.39$ &$14.18$ \\

   \multicolumn{1}{ c}{RS-ZF-dTHP-MMSEc} \rule{0pt}{3ex} &$30.60$ &$24.32$ &$18.06$ \\
   \hline\label{Error Var Performance Table}

\end{tabular}
\end{center}
\end{table}

\begin{figure}[h]
\begin{center}
\includegraphics[scale=0.45]{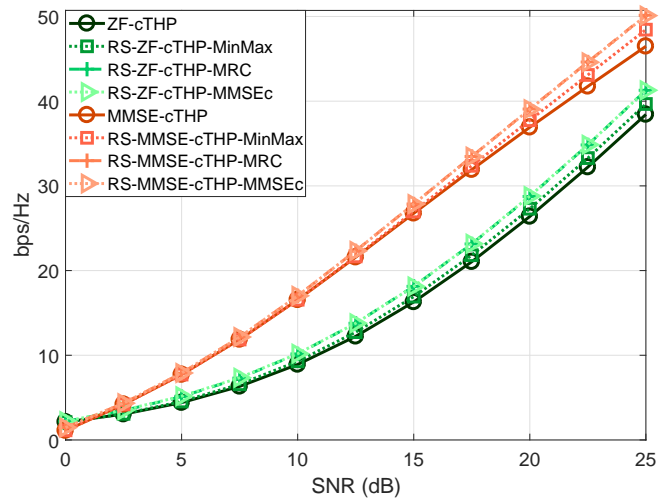}
\vspace{-1.0em}
\caption{{Sum-rate performance of a MU-MIMO system employing different THP precoding schemes with $\sigma_e^2=0.95(E_{tr})^{-0.6}$, $K=6$, $N_k=2$, $N_t=12$ and $\sigma_n^2=1$.}}
\label{Figure9}
\end{center}
\end{figure}

In the next example, we take into account that the quality of the CSIT improves as the SNR increases. We consider that the variance of the error scales as $\sigma_e^2=0.95(E_{tr})^{-0.6}$, which is inversely proportional to the SNR since the variance of the noise is fixed. Fig. \ref{Figure9} shows the results of the proposed strategies. Note that the curves show no saturation due the improvement in the CSIT quality. 
   
\begin{figure}[h]
\begin{center}
\includegraphics[scale=0.45]{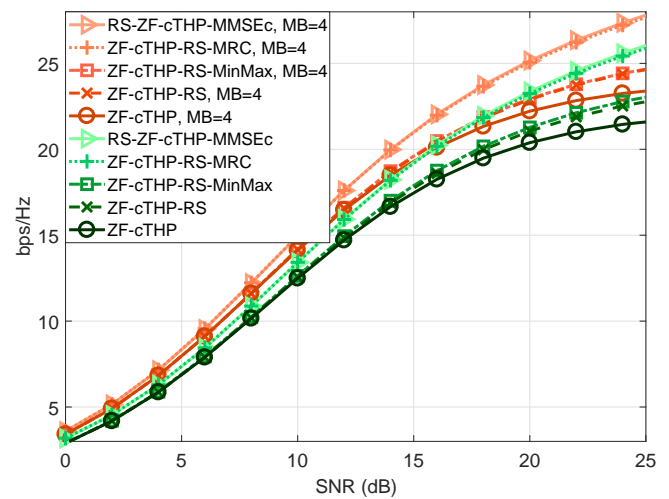} 
\vspace{-1.0em}
\caption{{Sum-rate performance of a MU-MIMO system employing different ZF-cTHP and MB-ZF-cTHP schemes with $\sigma_e^2=0.06$, $K=6$, $N_k=2$, $N_t=12$ and $\sigma_n^2=1$.}}
\label{Figure10}
\end{center}
\end{figure}

\begin{figure}[h]
\begin{center}
\includegraphics[scale=0.45]{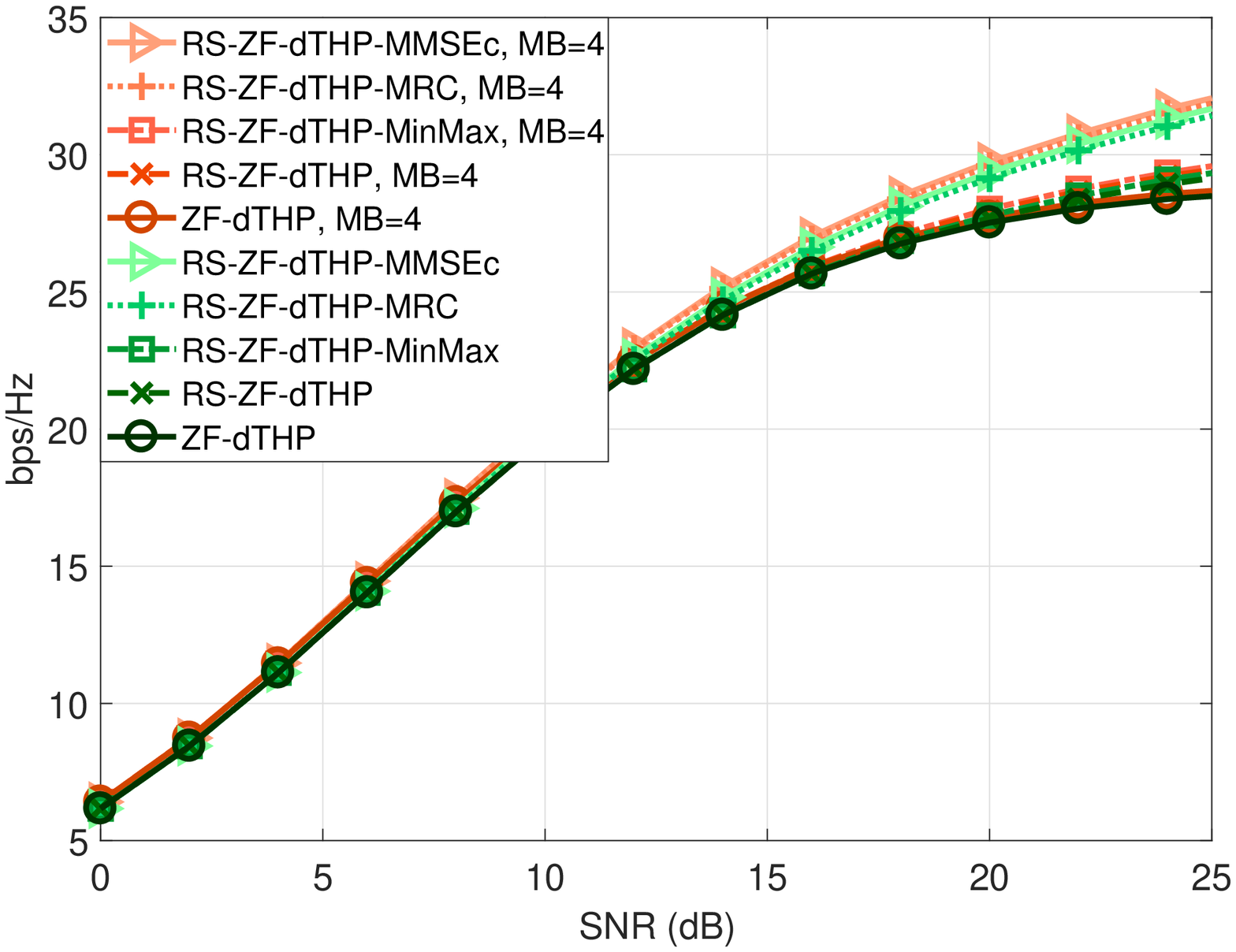}
\vspace{-1.0em}
\caption{{Sum-rate performance of a MU-MIMO system employing different ZF-dTHP and MB-ZF-dTHP schemes with $\sigma_e^2=0.06$, $K=6$, $N_k=2$, $N_t=12$ and $\sigma_n^2=1$.}}
\label{Figure11}
\end{center}
\end{figure}

We then consider the proposed RS-THP with the MB scheme. The term MB in the legends represents the number of branches employed in the simulation. We employ up to  four branches with both the ZF-cTHP and ZF-dTHP structures. First, power allocation of the common stream is carried out. Once the power allocation is performed, multiple THP branches are explored in order to select the branch that leads to the best performance.  Fig.  \ref{Figure10}
shows the result of the ZF-cTHP scheme. As expected, the best performance is obtained by the RS scheme with ZF-cTHP, four branches and the MMSEc combiner, denoted as RS-ZF-cTHP-MMSEc. {Moreover, the proposed RS-ZF-cTHP schemes with four branches outperform the conventional ZF-cTHP with four branches.} On the other hand, Fig. \ref{Figure11} summarizes the results for the ZF-dTHP structure. Once again {the proposed schemes attain a better performance than the conventional RS-ZF-dTHP with four branches.} The best performance is obtained by the RS-ZF-dTHP scheme with four branches and an MMSE stream combiner, denoted as RS-ZF-dTHP-MMSEc.

\begin{figure}[h]
\begin{center}
\includegraphics[scale=0.45]{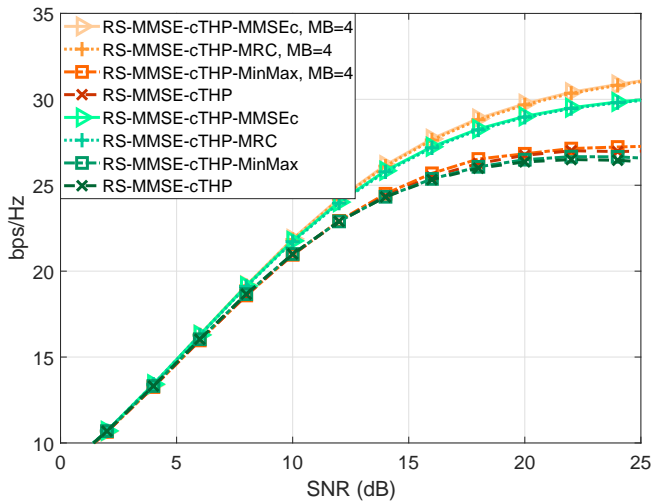}
\vspace{-1.0em}
\caption{{Sum-rate performance of a MU-MIMO system employing different MMSE-cTHP and MB-MMSE-cTHP schemes with $\sigma_e^2=0.06$, $K=6$, $N_k=2$, $N_t=12$ and $\sigma_n^2=1$.}}
\label{Figure12}
\end{center}
\end{figure}

In the last example, we asses the performance of the MMSE-THP precoder with MB. Once again, four branches were used to obtain the results shown in Fig. \ref{Figure12}. The RS-MMSEc-cTHP-MMSEc with four braches attains the best performance. As expected, the MB technique improves the results obtained since it considers four different symbol orderings. 

\section{Conclusion}
In this work, we have proposed RS-THP schemes along with stream combiners for multiple-antenna systems. Numerical results have shown that the non-linear structures proposed  attain higher sum-rate performance than conventional linear schemes under both, perfect and imperfect CSIT assumptions. RS-THP schemes along with the MMSEc combiner obtains the best sum-rate performance showing an increase of more than 20\% as compared to conventional MMSE-THP. This gain can be increased with the MB technique which considers different patterns to order the symbols as shown in the simulations where 4 different patterns were considered. Even this small number of branches produces a substantial gain which is up to 6\%. The proposed RS-THP schemes with common combiners exhibit greater benefits under imperfect CSIT. In those scenarios, RS-THP schemes increase considerably the robustness of systems, which demonstrates the efficiency of the RS scheme and the stream combiners against CSIT uncertainties.


%





\ifCLASSOPTIONcaptionsoff
  \newpage
\fi

\bibliographystyle{IEEEtran}
\bibliography{THPStreamComb} 

\begin{thebibliography}{10}
\providecommand{\url}[1]{#1}
\csname url@samestyle\endcsname
\providecommand{\newblock}{\relax}
\providecommand{\bibinfo}[2]{#2}
\providecommand{\BIBentrySTDinterwordspacing}{\spaceskip=0pt\relax}
\providecommand{\BIBentryALTinterwordstretchfactor}{4}
\providecommand{\BIBentryALTinterwordspacing}{\spaceskip=\fontdimen2\font plus
\BIBentryALTinterwordstretchfactor\fontdimen3\font minus
  \fontdimen4\font\relax}
\providecommand{\BIBforeignlanguage}[2]{{%
\expandafter\ifx\csname l@#1\endcsname\relax
\typeout{** WARNING: IEEEtran.bst: No hyphenation pattern has been}%
\typeout{** loaded for the language `#1'. Using the pattern for}%
\typeout{** the default language instead.}%
\else
\language=\csname l@#1\endcsname
\fi
#2}}
\providecommand{\BIBdecl}{\relax}
\BIBdecl

\bibitem{Li2010}
Q.~{Li}, G.~{Li}, W.~{Lee}, M.~{Lee}, D.~{Mazzarese}, B.~{Clerckx}, and
  Z.~{Li}, ``{MIMO} techniques in {WiMAX} and {LTE}: a feature overview,''
  \emph{IEEE Communications Magazine}, vol.~48, no.~5, pp. 86--92, May 2010.

\bibitem{Jones2015}
V.~{Jones} and H.~{Sampath}, ``Emerging technologies for {WLAN},'' \emph{IEEE
  Communications Magazine}, vol.~53, no.~3, pp. 141--149, Mar. 2015.

\bibitem{Parkvall2017}
S.~{Parkvall}, E.~{Dahlman}, A.~{Furuskar}, and M.~{Frenne}, ``{NR}: The new
  {5G} radio access technology,'' \emph{IEEE Communications Standards
  Magazine}, vol.~1, no.~4, pp. 24--30, Dec. 2017.

\bibitem{Shafi2017}
M.~{Shafi}, A.~F. {Molisch}, P.~J. {Smith}, T.~{Haustein}, P.~{Zhu}, P.~{De
  Silva}, F.~{Tufvesson}, A.~{Benjebbour}, and G.~{Wunder}, ``{5G}: A tutorial
  overview of standards, trials, challenges, deployment, and practice,''
  \emph{IEEE Journal on Selected Areas in Communications}, vol.~35, no.~6, pp.
  1201--1221, Jun. 2017.

\bibitem{mmimo}
R.~C. {de Lamare}, ``Massive mimo systems: Signal processing challenges and
  future trends,'' \emph{URSI Radio Science Bulletin}, vol. 2013, no. 347, pp.
  8--20, 2013.

\bibitem{wence}
W.~{Zhang}, H.~{Ren}, C.~{Pan}, M.~{Chen}, R.~C. {de Lamare}, B.~{Du}, and
  J.~{Dai}, ``Large-scale antenna systems with ul/dl hardware mismatch:
  Achievable rates analysis and calibration,'' \emph{IEEE Transactions on
  Communications}, vol.~63, no.~4, pp. 1216--1229, 2015.

\bibitem{mwc}
F.~L. {Duarte} and R.~C. {de Lamare}, ``Cloud-driven multi-way multiple-antenna
  relay systems: Joint detection, best-user-link selection and analysis,''
  \emph{IEEE Transactions on Communications}, vol.~68, no.~6, pp. 3342--3354,
  2020.

\bibitem{Joham2005}
M.~{Joham}, W.~{Utschick}, and J.~A. {Nossek}, ``Linear transmit processing in
  {MIMO} communications systems,'' \emph{IEEE Transactions on Signal
  Processing}, vol.~53, no.~8, pp. 2700--2712, Aug. 2005.

\bibitem{rmmseprec}
Y.~{Cai}, R.~C. {de Lamare}, L.~{Yang}, and M.~{Zhao}, ``Robust mmse precoding
  based on switched relaying and side information for multiuser mimo relay
  systems,'' \emph{IEEE Transactions on Vehicular Technology}, vol.~64, no.~12,
  pp. 5677--5687, 2015.

\bibitem{Spencer2004}
Q.~H. {Spencer}, A.~L. {Swindlehurst}, and M.~{Haardt}, ``Zero-forcing methods
  for downlink spatial multiplexing in multiuser {MIMO} channels,'' \emph{IEEE
  Transactions on Signal Processing}, vol.~52, no.~2, pp. 461--471, Feb. 2004.

\bibitem{Lai-UChoi2004}
{Lai-U Choi} and R.~D. {Murch}, ``A transmit preprocessing technique for
  multiuser {MIMO} systems using a decomposition approach,'' \emph{IEEE
  Transactions on Wireless Communications}, vol.~3, no.~1, pp. 20--24, Jan.
  2004.

\bibitem{lcbd}
K.~{Zu} and R.~C. d.~{Lamare}, ``Low-complexity lattice reduction-aided
  regularized block diagonalization for mu-mimo systems,'' \emph{IEEE
  Communications Letters}, vol.~16, no.~6, pp. 925--928, 2012.

\bibitem{Zu2013}
K.~{Zu}, R.~C. {de Lamare}, and M.~{Haardt}, ``Generalized design of
  low-complexity block diagonalization type precoding algorithms for multiuser
  {MIMO} systems,'' \emph{IEEE Transactions on Communications}, vol.~61,
  no.~10, pp. 4232--4242, Sep. 2013.

\bibitem{wlbd}
W.~{Zhang}, R.~C. {de Lamare}, C.~{Pan}, M.~{Chen}, J.~{Dai}, B.~{Wu}, and
  X.~{Bao}, ``Widely linear precoding for large-scale mimo with iqi: Algorithms
  and performance analysis,'' \emph{IEEE Transactions on Wireless
  Communications}, vol.~16, no.~5, pp. 3298--3312, 2017.

\bibitem{Stankovic2008}
V.~{Stankovic} and M.~{Haardt}, ``Generalized design of multi-user {MIMO}
  precoding matrices,'' \emph{IEEE Transactions on Wireless Communications},
  vol.~7, no.~3, pp. 953--961, Mar. 2008.

\bibitem{Sung2009}
H.~{Sung}, S.~. {Lee}, and I.~{Lee}, ``Generalized channel inversion methods
  for multiuser {MIMO} systems,'' \emph{IEEE Transactions on Communications},
  vol.~57, no.~11, pp. 3489--3499, Nov. 2009.

\bibitem{bbprec}
L.~T.~N. {Landau} and R.~C. {de Lamare}, ``Branch-and-bound precoding for
  multiuser mimo systems with 1-bit quantization,'' \emph{IEEE Wireless
  Communications Letters}, vol.~6, no.~6, pp. 770--773, 2017.

\bibitem{Tomlinson1971}
M.~{Tomlinson}, ``New automatic equaliser employing modulo arithmetic,''
  \emph{Electronics Letters}, vol.~7, no.~5, pp. 138--139, Mar. 1971.

\bibitem{Harashima1972}
H.~{Harashima} and H.~{Miyakawa}, ``Matched-transmission technique for channels
  with intersymbol interference,'' \emph{IEEE Transactions on Communications},
  vol.~20, no.~4, pp. 774--780, Aug. 1972.

\bibitem{Fischer2002}
R.~F.~H. {Fischer}, C.~{Windpassinger}, A.~{Lampe}, and J.~B. {Huber},
  ``Space-time transmission using {Tomlinson-Harashima} precoding,'' in
  \emph{2002 ITG Conf. on Source and Channel Coding}, 2002, pp. 139--147.

\bibitem{Windpassinger2004}
C.~{Windpassinger}, R.~F.~H. {Fischer}, T.~{Vencel}, and J.~B. {Huber},
  ``Precoding in multiantenna and multiuser communications,'' \emph{IEEE
  Transactions on Wireless Communications}, vol.~3, no.~4, pp. 1305--1316, Jul.
  2004.

\bibitem{Debels2018}
E.~{Debels} and M.~{Moeneclaey}, ``{SNR} maximization and modulo loss reduction
  for {Tomlinson-Harashima} precoding,'' \emph{EURASIP Journal on Wireless
  Communications and Networking}, vol. 2018, no. 257, Nov. 2018.

\bibitem{spa}
R.~C. {De Lamare} and R.~{Sampaio-Neto}, ``Minimum mean-squared error iterative
  successive parallel arbitrated decision feedback detectors for ds-cdma
  systems,'' \emph{IEEE Transactions on Communications}, vol.~56, no.~5, pp.
  778--789, 2008.

\bibitem{mfsic}
P.~{Li}, R.~C. {de Lamare}, and R.~{Fa}, ``Multiple feedback successive
  interference cancellation detection for multiuser mimo systems,'' \emph{IEEE
  Transactions on Wireless Communications}, vol.~10, no.~8, pp. 2434--2439,
  2011.

\bibitem{dfcc}
P.~{Li} and R.~C. {De Lamare}, ``Adaptive decision-feedback detection with
  constellation constraints for mimo systems,'' \emph{IEEE Transactions on
  Vehicular Technology}, vol.~61, no.~2, pp. 853--859, 2012.

\bibitem{tds}
P.~{Clarke} and R.~C. {de Lamare}, ``Transmit diversity and relay selection
  algorithms for multirelay cooperative mimo systems,'' \emph{IEEE Transactions
  on Vehicular Technology}, vol.~61, no.~3, pp. 1084--1098, 2012.

\bibitem{mbdf}
R.~C. {de Lamare}, ``Adaptive and iterative multi-branch mmse decision feedback
  detection algorithms for multi-antenna systems,'' \emph{IEEE Transactions on
  Wireless Communications}, vol.~12, no.~10, pp. 5294--5308, 2013.

\bibitem{1bitidd}
Z.~{Shao}, R.~C. {de Lamare}, and L.~T.~N. {Landau}, ``Iterative detection and
  decoding for large-scale multiple-antenna systems with 1-bit adcs,''
  \emph{IEEE Wireless Communications Letters}, vol.~7, no.~3, pp. 476--479,
  2018.

\bibitem{bfidd}
A.~G.~D. {Uchoa}, C.~T. {Healy}, and R.~C. {de Lamare}, ``Iterative detection
  and decoding algorithms for mimo systems in block-fading channels using ldpc
  codes,'' \emph{IEEE Transactions on Vehicular Technology}, vol.~65, no.~4,
  pp. 2735--2741, 2016.

\bibitem{listmtc}
R.~B. {Di Renna} and R.~C. {de Lamare}, ``Iterative list detection and decoding
  for massive machine-type communications,'' \emph{IEEE Transactions on
  Communications}, vol.~68, no.~10, pp. 6276--6288, 2020.

\bibitem{Wolniansky1998}
P.~W. {Wolniansky}, G.~J. {Foschini}, G.~D. {Golden}, and R.~A. {Valenzuela},
  ``{V-BLAST}: an architecture for realizing very high data rates over the
  rich-scattering wireless channel,'' in \emph{1998 URSI International
  Symposium on Signals, Systems, and Electronics}, Oct. 1998, pp. 295--300.

\bibitem{Wuebben2002}
D.~{Wübben}, J.~{Rinas}, R.~{Böhnke}, V.~{Kühn}, and K.~{Kammeyer},
  ``Efficient algorithm for detecting layered space-time codes,'' in \emph{2002
  ITG Conf. on Source and Channel Coding}, 2002, pp. 399--405.

\bibitem{Wubben2003}
D.~{Wubben}, R.~{Bohnke}, V.~{Kuhn}, and K.~. {Kammeyer}, ``{MMSE} extension of
  {V-BLAST} based on sorted {QR} decomposition,'' in \emph{2003 IEEE 58th
  Vehicular Technology Conference}, vol.~1, 2003, pp. 508--512.

\bibitem{Kusume2007}
K.~{Kusume}, M.~{Joham}, W.~{Utschick}, and G.~{Bauch}, ``Cholesky
  factorization with symmetric permutation applied to detecting and precoding
  spatially multiplexed data streams,'' \emph{IEEE Transactions on Signal
  Processing}, vol.~55, no.~6, pp. 3089--3103, Jun. 2007.

\bibitem{Fung2007}
C.~F. {Fung}, W.~{Yu}, and T.~J. {Lim}, ``Precoding for the multiantenna
  downlink: Multiuser {SNR} gap and optimal user ordering,'' \emph{IEEE
  Transactions on Communications}, vol.~55, no.~1, pp. 188--197, Jan. 2007.

\bibitem{Dao2010}
N.~{Dao} and Y.~{Sun}, ``User-selection algorithms for multiuser precoding,''
  \emph{IEEE Transactions on Vehicular Technology}, vol.~59, no.~7, pp.
  3617--3622, Sep. 2010.

\bibitem{Zu2014}
K.~{Zu}, R.~C. {de Lamare}, and M.~{Haardt}, ``Multi-branch
  {T}omlinson-{H}arashima precoding design for {MU-MIMO} systems: Theory and
  algorithms,'' \emph{IEEE Transactions on Communications}, vol.~62, no.~3, pp.
  939--951, Mar. 2014.

\bibitem{Sun2015}
L.~{Sun}, J.~{Wang}, and V.~C.~M. {Leung}, ``Joint transceiver, data streams,
  and user ordering optimization for nonlinear multiuser {MIMO} systems,''
  \emph{IEEE Transactions on Communications}, vol.~63, no.~10, pp. 3686--3701,
  Oct. 2015.

\bibitem{locsme}
H.~{Ruan} and R.~C. {de Lamare}, ``Robust adaptive beamforming using a
  low-complexity shrinkage-based mismatch estimation algorithm,'' \emph{IEEE
  Signal Processing Letters}, vol.~21, no.~1, pp. 60--64, 2014.

\bibitem{okspme}
------, ``Robust adaptive beamforming based on low-rank and cross-correlation
  techniques,'' \emph{IEEE Transactions on Signal Processing}, vol.~64, no.~15,
  pp. 3919--3932, 2016.

\bibitem{Clerckx2016}
B.~{Clerckx}, H.~{Joudeh}, C.~{Hao}, M.~{Dai}, and B.~{Rassouli}, ``Rate
  splitting for {MIMO} wireless networks: a promising {PHY}-layer strategy for
  {LTE} evolution,'' \emph{IEEE Communications Magazine}, vol.~54, no.~5, pp.
  98--105, May 2016.

\bibitem{Piovano2017}
E.~{Piovano} and B.~{Clerckx}, ``Optimal {DoF} region of the $k$ -user {MISO}
  {BC} with partial {CSIT},'' \emph{IEEE Communications Letters}, vol.~21,
  no.~11, pp. 2368--2371, Nov. 2017.

\bibitem{Mao2018}
Y.~{Mao}, B.~{Clerckx}, and V.~{Li}, ``Rate-splitting multiple access for
  downlink communication systems: Bridging, generalizing and outperforming
  {SDMA} and {NOMA},'' \emph{EURASIP Journal on Wireless Communications and
  Networking}, vol. 2018, no.~1, p. 133, May 2018.

\bibitem{Clerckx2020}
B.~{Clerckx}, Y.~{Mao}, R.~{Schober}, and H.~V. {Poor}, ``Rate-splitting
  unifying {SDMA}, {OMA}, {NOMA}, and multicasting in {MISO} broadcast channel:
  A simple two-user rate analysis,'' \emph{IEEE Wireless Communications
  Letters}, vol.~9, no.~3, pp. 349--353, Mar. 2020.

\bibitem{Mao2020}
Y.~{Mao} and B.~{Clerckx}, ``Beyond dirty paper coding for multi-antenna
  broadcast channel with partial {CSIT}: A rate-splitting approach,'' \emph{in
  submission to IEEE Transactions on Communications}, 2020.

\bibitem{Joudeh2016}
H.~{Joudeh} and B.~{Clerckx}, ``Sum-rate maximization for linearly precoded
  downlink multiuser {MISO} systems with partial {CSIT}: A rate-splitting
  approach,'' \emph{IEEE Transactions on Communications}, vol.~64, no.~11, pp.
  4847--4861, Nov. 2016.

\bibitem{Hao2015}
C.~{Hao}, Y.~{Wu}, and B.~{Clerckx}, ``Rate analysis of two-receiver {MISO}
  broadcast channel with finite rate feedback: A rate-splitting approach,''
  \emph{IEEE Transactions on Communications}, vol.~63, no.~9, pp. 3232--3246,
  Sep. 2015.

\bibitem{Flores2018}
A.~R. {Flores}, B.~{Clerckx}, and R.~C. {de Lamare}, ``Tomlinson-harashima
  precoded rate-splitting for multiuser multiple-antenna systems,'' in
  \emph{2018 15th International Symposium on Wireless Communication Systems
  (ISWCS)}, Lisbon, Aug. 2018, pp. 1--6.

\bibitem{Joudeh2017}
H.~{Joudeh} and B.~{Clerckx}, ``Rate-splitting for max-min fair multigroup
  multicast beamforming in overloaded systems,'' \emph{IEEE Transactions on
  Wireless Communications}, vol.~16, no.~11, pp. 7276--7289, Nov. 2017.

\bibitem{Hao2017}
C.~{Hao}, B.~{Rassouli}, and B.~{Clerckx}, ``Achievable {DoF} regions of {MIMO}
  networks with imperfect {CSIT},'' \emph{IEEE Transactions on Information
  Theory}, vol.~63, no.~10, pp. 6587--6606, Oct. 2017.

\bibitem{Lu2018}
G.~{Lu}, L.~{Li}, H.~{Tian}, and F.~{Qian}, ``{MMSE}-based precoding for rate
  splitting systems with finite feedback,'' \emph{IEEE Communications Letters},
  vol.~22, no.~3, pp. 642--645, Mar. 2018.

\bibitem{Flores2020}
A.~R. {Flores}, R.~C. {De Lamare}, and B.~{Clerckx}, ``Linear precoding and
  stream combining for rate splitting in multiuser {MIMO} systems,'' \emph{IEEE
  Communications Letters}, vol.~24, no.~4, pp. 890--894, Jan. 2020.

\bibitem{jidf}
R.~C. {de Lamare} and R.~{Sampaio-Neto}, ``Adaptive reduced-rank processing
  based on joint and iterative interpolation, decimation, and filtering,''
  \emph{IEEE Transactions on Signal Processing}, vol.~57, no.~7, pp.
  2503--2514, 2009.

\bibitem{Vu2007}
M.~{Vu} and A.~{Paulraj}, ``Mimo wireless linear precoding,'' \emph{IEEE Signal
  Processing Magazine}, vol.~24, no.~5, pp. 86--105, Sep. 2007.

\bibitem{Love2008}
D.~J. {Love}, R.~W. {Heath}, V.~K. {N. Lau}, D.~{Gesbert}, B.~D. {Rao}, and
  M.~{Andrews}, ``An overview of limited feedback in wireless communication
  systems,'' \emph{IEEE Journal on Selected Areas in Communications}, vol.~26,
  no.~8, pp. 1341--1365, Oct. 2008.

\bibitem{Yang2013}
S.~{Yang}, M.~{Kobayashi}, D.~{Gesbert}, and X.~{Yi}, ``Degrees of freedom of
  time correlated {MISO} broadcast channel with delayed {CSIT},'' \emph{IEEE
  Transactions on Information Theory}, vol.~59, no.~1, pp. 315--328, Jan. 2013.

\bibitem{WeiYu2005}
{Wei Yu}, D.~P. {Varodayan}, and J.~M. {Cioffi}, ``Trellis and convolutional
  precoding for transmitter-based interference presubtraction,'' \emph{IEEE
  Transactions on Communications}, vol.~53, no.~7, pp. 1220--1230, Jul. 2005.

\bibitem{jpba}
Y.~{Jiang}, Y.~{Zou}, H.~{Guo}, T.~A. {Tsiftsis}, M.~R. {Bhatnagar}, R.~C. {de
  Lamare}, and Y.~{Yao}, ``Joint power and bandwidth allocation for
  energy-efficient heterogeneous cellular networks,'' \emph{IEEE Transactions
  on Communications}, vol.~67, no.~9, pp. 6168--6178, 2019.

\bibitem{DeLamare2008}
R.~C. {De Lamare} and R.~{Sampaio-Neto}, ``Minimum mean-squared error iterative
  successive parallel arbitrated decision feedback detectors for {DS-CDMA}
  systems,'' \emph{IEEE Transactions on Communications}, vol.~56, no.~5, pp.
  778--789, May 2008.

\bibitem{Ahmad2019}
A.~A. {Ahmad}, J.~{Kakar}, H.~{Dahrouj}, A.~{Chaaban}, K.~{Shen}, A.~{Sezgin},
  T.~Y. {Al-Naffouri}, and M.~{Alouini}, ``Rate splitting and common message
  decoding for {MIMO} {C-RAN} systems,'' in \emph{2019 IEEE 20th International
  Workshop on Signal Processing Advances in Wireless Communications (SPAWC)},
  2019.

\bibitem{Kolawole2018}
O.~{Kolawole}, A.~{Panazafeironoulos}, and T.~{Ratnarajah}, ``A rate-splitting
  strategy for multi-user millimeter-wave systems with imperfect {CSI},'' in
  \emph{2018 IEEE 19th International Workshop on Signal Processing Advances in
  Wireless Communications (SPAWC)}, 2018.

\bibitem{Golub1996}
G.~{Golub} and C.~{Van Loan}, \emph{Matrix Computations}.\hskip 1em plus 0.5em
  minus 0.4em\relax The Johns Hopkins University Press, 1996.

\end{thebibliography}
\end{document}